\documentclass[longauthor, times]{aastex61}

\received{\today}
\revised{}
\accepted{}
\submitjournal{ApJ}

\usepackage{color,comment}
\usepackage[normalem]{ulem}

\newcommand{\red}[1]{{\textcolor{red}{#1}}}

\newcommand{\angstrom}{\text{\normalfont\AA}}

\shorttitle{Dust Attenuation Laws in Galaxy Simulations}
\shortauthors{Narayanan, Conroy, Dav\'e, Johnson \& Popping}

\begin{document}

\title{A Theory for the Variation of Dust Attenuation Laws in Galaxies}

\correspondingauthor{Desika Narayanan}
\email{desika.narayanan@ufl.edu}

\author[0000-0002-7064-4309]{Desika Narayanan}
\affil{Department of Astronomy, University of Florida, 211 Bryant Space Sciences Center, Gainesville, FL 32611 USA}
\affil{University of Florida Informatics Institute, 432 Newell Drive, CISE Bldg E251, Gainesville, FL 32611}
\affil{Cosmic Dawn Center (DAWN), Niels Bohr Institute, University of Copenhagen, Juliane Maries vej 30, DK-2100 Copenhagen, Denmark}
\author[0000-0002-1590-8551]{Charlie Conroy}
\affil{Department of Astronomy, Harvard University, 60 Garden Street, Cambridge, MA 02138}
\author[0000-0003-2842-9434]{Romeel Dav\'e}
\affil{Institute for Astronomy, Royal Observatory Edinburgh, EH9 3HJ, UK}
\affil{University of the Western Cape, Bellville, Cape Town 7535, South Africa}
\affil{South African Astronomical Observatory, Cape Town 7925, South Africa}
\author[0000-0002-9280-7594]{Benjamin D. Johnson}
\affil{Department of Astronomy, Harvard University, 60 Garden Street, Cambridge, MA 0213}
\author[0000-0003-1151-4659]{Gerg\"o Popping}
\affil{Max-Planck-Institut f\"ur Astronomie, K\"onigstuhl 17, D-69117 Heidelberg, Germany}



\begin{abstract}
 In this paper, we provide a physical model for the origin of
 variations in the shapes and bump strengths of dust attenuation laws
 in galaxies by combining a large suite of cosmological ``zoom-in''
 galaxy formation simulations with 3D Monte Carlo dust radiative
 transfer calculations.  We model galaxies over $3$ orders of
 magnitude in stellar mass, ranging from Milky Way like systems
 through massive galaxies at high-redshift.  Critically, for these
 calculations we employ a constant underlying dust extinction law in
 all cases, and examine how the role of geometry and radiative
 transfer effects impact the resultant attenuation curves.  Our main
 results follow.  Despite our usage of a constant dust extinction
 curve, we find dramatic variations in the derived attenuation laws.
 The slopes of normalized attenuation laws depend primarily on the
 complexities of star-dust geometry.  Increasing fractions of
 unobscured young stars flatten normalized curves, while increasing
 fractions of unobscured old stars steepen curves.  Similar to the
 slopes of our model attenuation laws, we find dramatic variation in
 the $2175 \angstrom$ ultraviolet (UV) bump strength, including a
 subset of curves with little to no bump.  These bump strengths are
 primarily influenced by the fraction of unobscured O and B stars in
 our model, with the impact of scattered light having only a secondary
 effect.  Taken together, these results lead to a natural relationship
 between the attenuation curve slope and $2175 \angstrom$ bump
 strength.  Finally, we apply these results to a $25$ Mpc/$h$ box
 cosmological hydrodynamic simulation in order to model the expected
 dispersion in attenuation laws at integer redshifts from $z=0-6$.  A
 significant dispersion is expected at low redshifts, and decreases
 toward $z=6$.  We provide tabulated results for the best fit median
 attenuation curve at all redshifts.

\end{abstract}




\section{Introduction} \label{section:introduction}

Astrophysical dust is pervasive in the interstellar medium (ISM) of
most galaxies. This dust can both absorb stellar radiation, as well as
scatter light into and out of the line of sight \citep{witt00a}.
Physical properties of galaxies that derive from optical and
ultraviolet (UV) radiation, such as the star formation rate (SFR),
stellar mass ($M_*$), and stellar ages must therefore account for
these dust-driven effects. Unfortunately, how exactly to do this is
complicated \citep[see reviews by][]{walcher11a,conroy13a}.
Beyond the aforementioned effects of extinction and scattering,
  the spatial distribution between stars of different ages and the
  dusty ISM (hereafter simply referred to as ``the geometry'') can
dramatically impact dust attenuation\footnote{Following typical
  convention, we define the following.  {\it Extinction} measures the
  loss of photons along the line of sight due to absorption, as well
  as scattering out of the line of sight.  {\it Attenuation} refers to
  extinction, but also taking into account scattering back into the
  line of sight, as well as the geometric distribution of dust with
  respect to stars} laws in galaxies. These effects are typically
described via an attenuation curve, which quantifies the optical depth
($\tau$), as either a function of wavelength ($\lambda$), or $x\equiv
1/\lambda$.

\begin{figure*}
\begin{tabular}{cc}
\hspace{-0.25cm}
\includegraphics[]{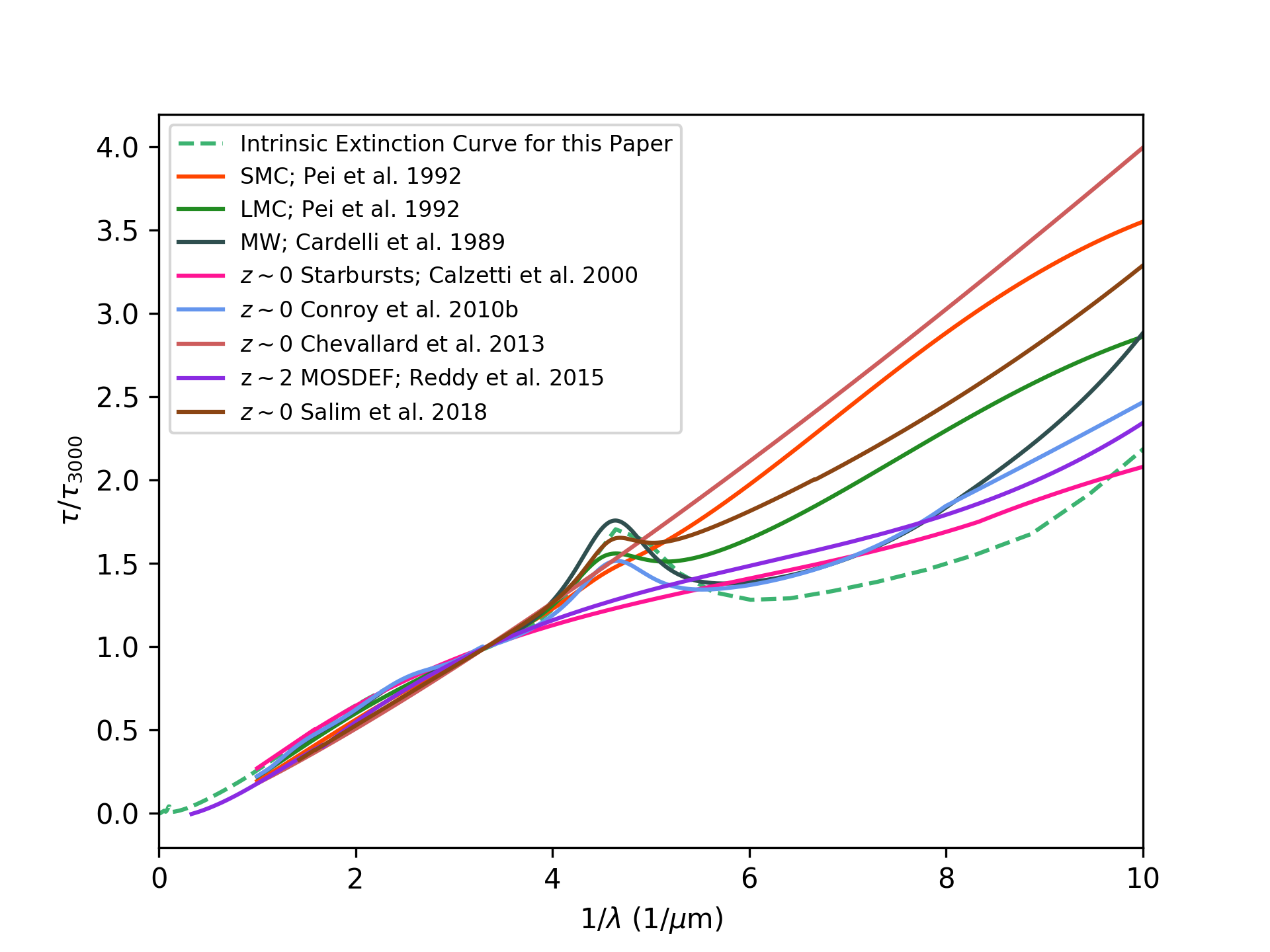}
\end{tabular}
    \caption{Compilation of selected extinction attenuation curve
      constraints taken from the literature. The extinction curves are
      the \citet{cardelli89a} and \citet{pei92a} curves, while the
      remainder are attenuation curves. The solid lines show reference
      curves from the literature, while the green dashed line shows
      the intrinsic extinction curve used in our cosmological galaxy
      formation simulations in this paper \citep{weingartner01a}.  \label{figure:literature}}
\end{figure*}

Decades of constraints of observed extinction attenuation curves have
evidenced a number of clear features \citep[see reviews
  by][]{calzetti01a,draine03a,galliano17a}.  First, there is a steep
rise toward the ultraviolet due to absorption by small grains such
that shorter wavelength photons are preferentially removed from the
line of sight.  In the near ultraviolet (NUV), there is (sometimes) an
absorption bump near $2175 \angstrom$, first reported by
\citet{stecher65a}, that is potentially associated with polycyclic
aromatic hydrocarbons \citep[e.g.][]{weingartner01a}, though
some models suggest a graphite-based origin
\citep[e.g.][]{stecher65b,hoyle62a,draine93a}, or other species yet
\citep[e.g.][]{wada99a}.  This $2175 \angstrom$ UV bump is known to
vary dramatically in strength, a topic we return to later.  There is
some evidence for an optical knee due to scattering by large grains
\citep{galliano17a}, and the extinction curve takes a powerlaw shape
at longer NIR wavelengths.   In Figure~\ref{figure:literature}, we show
a compilation of a number of literature extinction and attenuation
curves  that will serve as a useful reference throughout this paper.

These observations have revealed a diversity in derived extinction
curves in the Milky Way and nearby galaxies.  For example, when
parameterizing the inverse of the slope of the extinction curve in the
optical as $R_{\rm V} \equiv A_{\rm V}/\left(A_{\rm B}-A_{\rm
  V}\right)$, observations have ranged from values of $R_{\rm V}$ as
low as $\sim 2$ \citep{welty92a} to nearly $R_{\rm V} \sim 6$
\citep{cardelli89a,fitzpatrick99a}.  Similarly, \citet{fitzpatrick90a}
demonstrated a broad range of extinction curves amongst different
Milky Way sightlines via International Ultraviolet Explorer (IUE)
observations.  \citet{cardelli89a} and \citet{fitzpatrick90a} showed
that variations in both the attenuation curve shape and $2175
\angstrom $ UV bump strength can be well-parameterized by the total to
selective extinction $R_{\rm V}$.  The sightline average attenuation
curve of the Milky Way has $R_{\rm V} \approx 3.1$
\citep[e.g.][]{rieke85a}.  Outside of our galaxy, extinction curves
have been measured using this method for the Magallanic clouds, and
Andromeda.  \citet{pei92a} and \citet{gordon03a} derived attenuation
curves for the Magallanic clouds, and showed that the Small Magallanic
Cloud (SMC) bar, for example, has a steeper curve than the Large
Magallanic Cloud (LMC) average when normalizing at optical
wavelengths.  On average, both of the Magallanic clouds have lower
$R_{\rm V}$ ratios than the Milky Way, which may be due to their lower
metallicities \citep{misselt99a,calzetti01a}.  At the same time,
\citet{clayton15a} found that M31 has a relatively similar average
curve as the Milky Way.

At larger distances, emission from unresolved stellar populations are
measured in aggregate (as opposed to in individual stellar pairs), and
  the role of star-dust geometry can introduce significant
complications.  In this regime, constraints on {\it extinction} curves
become impossible, and instead measurements are made of {\it
  attenuation} curves.  When measuring attenuation toward stars in
unresolved galaxies (i.e. $A_{\rm star}$), methods include utilizing
the ratio of the infrared excess (itself a ratio between the infrared
luminosity to a monochromatic UV luminosity) to the UV slope
\citep[i.e. the IRX-$\beta$
  relation;][]{meurer99a,johnson07b,casey14b}, color-magnitude diagram
fitting \citep[e.g.][]{dalcanton15a}, and SED fitting techniques
\citep[e.g.][]{papovich01a,buat11a,kriek13a,conroy13a,leja17a,salim18a}.

Akin to extinction curves within the Milky Way, attenuation curves in
galaxies near and far have shown a dramatic range in observed shapes.
In a series of papers,
\citet{calzetti94a,kinney94a,calzetti97a,boker99a} and
\citet{calzetti00a} investigated attenuation laws toward both stars
and HII regions in local Universe starburst galaxies.  These
researchers found that these galaxies, on average, have greyer (also
described as 'flatter' or 'shallower' throughout this paper)
optical/UV attenuation slopes (when normalized in the optical) than
the Milky Way, with an average $R_{\rm V} = 4.05$.  \citet{johnson07a}
derive a mean attenuation law in a sample of $\sim 1000$ galaxies of
$A_{\lambda} \sim \lambda^{-0.7}$, with little variation when binned
by stellar mass.  \citet{wild11a} analyzed $\sim 23000$ galaxies from
the Sloan Digital Sky Survey, and found that the slope of attenuation
curves varies strongly with galaxy axial ratio, though only weakly
with their specific star formation rate.  The slopes of these curves
vary with stellar mass surface density, with higher stellar mass
surface densities correlating with steeper curves.  Meanwhile,
\citet{battisti16a} found, for $\sim 10000$ star-forming galaxies at
$z<0.1$, an average attenuation curve similar to that derived by
\citet{calzetti00a}, with an observed range comparable to the factor
$\sim 2$ dispersion in $\tau_{\rm FUV}/\tau_{\rm V}$ found by
\citet{wild11a}.  \citet{battisti17a} expanded on this study, and
found little correlation between these curves and a range of galaxy
physical properties, including mean stellar age, the specific star
formation rate, stellar mass and metallicity.  The consensus from
local galaxy observations is that, aside from potentially stellar mass
surface density \citep[e.g.][]{wild11a}, there appear to be relatively
few correlations between the slope of dust attenuation curves and
galaxy physical property. 

At high-redshift, the most commonly assumed attenuation curve is a
\citet{calzetti00a} law.  Owing to the sensitivity of derived physical
parameters on the assumed attenuation law in SED modeling
\citep[e.g.][]{salim07a}, a number of authors have attempted to
constrain observed attenuation laws at high-redshift to assess the
validity of the typical assumption of a \citet{calzetti00a} relation.
For example, \citet{reddy04a} compared X-ray and radio
(i.e. relatively extinction-free) derived SFRs with that derived from
Calzetti-corrected UV photometry of UV-selected star forming galaxies
between $1.5 \la z \la 3$ in GOODS-N, and found consistent results.
This implies, on average, that an attenuation curve similar to that
derived by \citet{calzetti00a} is a reasonable descriptor of the
attenuation properties of these galaxies.  A number of other works,
using complementary methods, have found reasonable consistency with a
\citet{calzetti00a} attenuation curve at high-redshift
\citep[e.g.][]{scoville15a,cullen18a,cullen17a}.  Much of this work
has focused on the location of galaxies at high-redshift on the
IRX-$\beta$ plane, and their consistency with a \citet{calzetti00a}
attenuation curve 
\citep[e.g.][]{seibert02a,reddy08a,pannella09a,siana09a,reddy12a,
  heinis13a,to14a,bourne16a,mclure17a}.

  At the same time, \citet{reddy15a} derived attenuation curves for
  $\sim 200$ $z\sim 2$ galaxies from the MOSFIRE Deep Evolution Field
  (MOSDEF) survey, and found evidence for a composite attenuation
  curve that is similar to the \citet{calzetti00a} form at short
  wavelengths, and the SMC extinction curve at $\lambda \ga 2500
  \angstrom$.  Similarly, \citet{lofaro17a} find evidence for a
  composite attenuation curve in a sample of $z \sim 2$ galaxies
  selected for their infrared luminosity.  \citet{salmon16a} found a
  diverse range of attenuation laws in $z \sim 1.5-3$ galaxies from
  the CANDELS survey with little correlation with galaxy physical
  property, and \citet{shivaei15a} required curves steeper than
  \citet{calzetti00a} to match the observed IR/UV ratios in $\sim 250
  \ z \sim 2$ star-forming galaxies.  And, in the same vein,
    dramatic outliers from the \citet{meurer99a} IRX-$\beta$ relation
    at high-redshift have pointed to underlying variations in
    attenuation laws.  This said, other complicating factors may
    contribute to deviations from the locally calibrated
    \citet{meurer99a} relation
    \citep[e.g.][]{bell02a,kong04a,grasha13a,koprowski16a,safarzadeh17a,popping17a,narayanan18a}.

Similar to variations in the attenuation curve slope, galaxies in both
the local Universe and at high-redshift exhibit a wide range of $2175
\angstrom$ UV absorption bump strengths.  Both the SMC and average
attenuation curve for nearby starburst galaxies lack strong bump
features \citep{gordon00a,calzetti00a}, and \citet{gordon99a} find
relatively small bump strengths in attenuation curves for a sample of
high-redshift galaxies from the Hubble Deep Fields.  Meanwhile,
a number of studies at both low redshift
\citep{conroy10c,wild11a,battisti16a,salim18a}, and high-$z$
\citep{motta02a,burgarella05a,york06a,stratta07a,noll07a,noll09a,eliasdottir09a,buat11a,scoville15a}
have found evidence of $2175 \angstrom$ absorption bumps of varying
strengths.  A substantial step forward was made by \citet{kriek13a},
who not only demonstrated evidence of a bump in the attenuation curves
of $z \sim 2$ star-forming galaxies, but also that the bump strength
varies with the slope of the attenuation curve (such that the steepest
curves have the strongest bumps).

Given the strong dependence on inferred galaxy physical properties on
attenuation curves, understanding the origin of their shape variations
is critical.  Here, modeling can provide significant insight as to the
physical drivers of variations in both attenuation law slopes and
feature strengths. Models designed to understand attenuation laws in
galaxies generally fall into one of three methodology camps.  The core
of each involves radiative transfer modeling of stellar light through
a dusty interstellar medium, and the primary difference lies in how
the geometry of stars and the dust is modeled.  The general trade off
is that increases in model complexity are typically associated with
increased astrophysical realism, though decreased ability to develop
controlled numerical experiments.

The most simplified models typically involve illuminating a slab or
shell-like geometry with a stellar radiation source.  Early works
include \citet{witt96a,witt00a}, who used these models to demonstrate
that increasingly mixed star-dust geometries result in flatter
(grayer) attenuation curves.  \citet{seon16a} expanded upon these
significantly by developing a model for a turbulent ISM with a
lognormal density profile and a clumpy medium to investigate the origin
of attenuation law shape and bump strength variations.  While these
models can neatly isolate individual physical effects, they do not
model galaxies as a whole, taking into account the radiative transfer
from  stars with a distribution of stellar ages and metallicities
through a cosmologically evolved interstellar medium.

The latter two modeling categories analyze attenuation laws for
galaxies as a whole.  These include hydrodynamic simulations of
idealized galaxies evolving without a cosmological context, and
simulations that employ semi-analytic modeling techniques within a
cosmological framework.  In the former category,
\citet{jonsson06a,rocha08a,natale15a,hayward15a} and \citet{hou17a}
coupled idealized models of isolated disk galaxies and
galaxy mergers with 3D dust radiative transfer to model their dust
attenuation properties. Such simulations can achieve very high
resolution, but do not include the hierarchical growth of galaxies
that can result in more diverse (and realistic) morphologies
\citep[e.g.][]{abruzzo18a} leading to variations in dust
attenuation. This is especially true at early epochs when dusty disks
are thicker and clumpier, and are responding dynamically to high rates
of gas inflows and outflows.  Meanwhile, semi-analytic models (SAMs)
have been developed to model dust attenuation in a cosmological
context
\citep[e.g.][]{granato00a,fontanot09a,fontanot11a,gonzalezperez13a,wilkins12a}.
Such models do not track baryonic growth directly, but rather track
dark matter growth and make physically-motivated but simplistic
assumptions about the resulting galaxy properties. The benefit is
that, when coupled with spectrophotometric dust radiative transfer
calculations, SAMs are able to efficiently produce cosmological
volumes of galaxies with SED information. But since the baryonic
matter is characterized through analytic expressions
\citep[e.g.][]{popescu11a}, the geometry of the ISM and stars is not
directly predicted but rather based on simplified assumptions.

What is missing thus far is a theoretical interpretation of dust
attenuation laws in the context of models that directly track the
hydrodynamic growth and evolution of galaxies from cosmological
conditions with sufficient resolution to model the impact of complex
star-dust geometries.  Cosmological hydrodynamic simulations are
attractive in their ability to directly track the hierarchical growth
of galaxies, and therefore provide relatively realistic star-dust
geometries for stellar populations with a distribution of
metallicities and formation ages.  In this paper, we present the first
ever theoretical model for dust attenuation curves in galaxies from
cosmological hydrodynamic simulations.

To do this, we will employ the cosmological ``zoom'' technique, where
we focus on individual halos (and their associated baryons) in
cosmological simulations at very high resolution.  We couple these
simulations with 3-dimensional dust radiative transfer to model how
the intrinsic stellar light is absorbed and scattered, and
consequently develop a model for dust attenuation laws in galaxies.
Critical to the interpretation of our results: in this model, we will
assume an underlying dust extinction law (i.e. we will hold the
properties of dust grains fixed), and ask how geometry and radiative
transfer effects drive variations in the attenuation law slope and UV
bump strength. In \S~\ref{section:simulations} we describe our
simulation setup; in \S~\ref{section:sps_models} we build physical
insight via simplified population synthesis models; in
\S~\ref{section:results}, we apply this insight to direct results from
cosmological zoom galaxy formation simulations.  Here, we explore the
origin of variations in the slope and the bump strength in attenuation
laws in galaxies.  In \S~\ref{section:discussion} we provide discussion,
and we summarize in \S~\ref{section:summary}.  Throughout, we assume a
cosmology of $(\Omega_0,\Omega_\Lambda,\Omega_{\rm b},h$) =
$(0.3,0.7,0.48,0.68)$.

\section{Simulation Methodology}
\label{section:simulations}
\subsection{Overview}
In order to model dust attenuation curves, we must know both the
intrinsic stellar spectrum in model galaxies, as well as the observed
SED.  To generate these, we first simulate a series of model galaxies
using the cosmological zoom technique.  We then determine the stellar
SEDs for these model galaxies based on the individual metallicities
and ages for the star particles via {\sc fsps}  calculations.  These
stellar SEDs are propagated through the dusty interstellar medium via
{\sc hyperion} dust radiative transfer (wrapped in the {\sc powderday}
code package), and compared to the final observed attenuated UV-optical
SED in order to determine the attenuation curve.  The underlying {\it
  extinction} properties of the grains are fixed throughout, and we
utilize the aforementioned simulations to understand the impact of
dust geometry and radiative transfer effects on the observed {\it
  attenuation} curves.

\subsection{Cosmological Zoom Galaxy Formation Simulations}

Our model galaxy suite is generated from the {\sc mufasa}-zoom
simulation series, which are zoomed galaxies from the {\sc mufasa}
cosmological simulation \citep{dave16a,dave17a,dave17b}.  This zoom
simulation methodology has been described in detail in
\citet{olsen17a}, \citet{narayanan18a}, \citet{abruzzo18a}, and
\citet{privon18a}.  We therefore refer the reader to those works for
details, and summarize the salient points here.

All of our simulations are conducted with a modified version of the
{\sc gizmo} hydrodynamic code \citep{hopkins15a,hopkins17a}, which
builds off of the code base in {\sc gadget-3} \citep{springel05a}.  We
first simulate a coarse resolution dark matter only run from $z=249$
down to $z=0$ using initial conditions generated from {\sc music}
\citep{hahn11a}.  The initial conditions for this dark matter box are
the exact same as those used in the {\sc mufasa} cosmological
hydrodynamic simulation.  This coarse resolution run is done in a $50
h^{-1}$ Mpc volume, and includes $512^3$ dark matter particles,
resulting in dark matter mass resolution of $7.8 \times 10^{8} h^{-1}
M_\odot$.  It is from this dark matter only simulation that we select
our model halos to re-simulate at much higher resolution (and with
baryons included).

We re-simulate eight halos over a broad mass range to final redshifts
$z_{\rm final}$, where $z_{\rm final}$ varies based on the halo mass.
Five of these halos are described in \citet{narayanan18a}, and we have
include three additional models in this paper.  In
Table~\ref{table:ics}, we describe their physical
properties\footnote{We order the simulations in Table~\ref{table:ics}
  by their $z=2$ halo mass.  In the case of halo $0$, which did not
  reach $z=2$, we order in terms of its expected $z=2$ mass from the
  parent dark matter cosmological simulation.}, and present their
location in stellar mass-halo mass space and SFR-$M_*$ space in
\citet{abruzzo18a}.  These simulations span roughly three decades
in halo mass, and represent galaxies ranging from Milky Way mass (at
$z=0$) through halo masses comparable to luminous dusty galaxies at $z
\sim 2$ \citep[e.g.][]{narayanan15a,casey14a}.

We identify halos using {\sc caesar} \citep{thompson15a}.  For each halo of
interest for resimulation, we build an ellipsoidal mask around all
particles encapsulating a radius $2.5 \times$ the distance of the
farthest particle from halo center, and define this as the Lagrangian
region for resimulation.  We then track these particles back to
$z=249$, split these to the desired resolution, and re-run the
simulation to $z=z_{\rm final}$ with hydrodynamics turned on.  We have
zero low-resolution particles contaminating any halo presented here
within three virial radii.

Our simulations use the suite of physics developed for the {\sc
  mufasa} cosmological hydrodynamic simulations
\citep{dave16a,dave17a,dave17b}.  Here, stars form in dense molecular
gas, and the H$_2$ fraction is calculated using the
\citet{krumholz09a} methodology, tying the molecular fraction to the
gas surface density and metallicity \citep{thompson14a}.  We
impose a minimum metallicity for star formation of $Z=10^{-3}
Z_\odot$.  This star formation occurs at a rate following a volumetric
\citet{schmidt59a} relation, with an imposed efficiency per free fall
time of $\epsilon_* = 0.02$, as motivated by observations
\citep{kennicutt98a,kennicutt12a,narayanan08b,narayanan12a,hopkins13a}.

Feedback from massive stars are modeled via the {\sc mufasa} decoupled
two-phase wind scheme.  The wind physics are presented in detail in
\citet{dave16a}, and we refer the reader to that work for a full
description of the wind model.  Briefly, the modeled stellar winds
have a probability for ejection that is modeled as a fraction of the
SFR probability.  This fraction is derived from the best-fitting
relation from the Feedback in Realistic Environments
\citep{muratov15a,hopkins14b,hopkins17b} simulation suite.  The
dependence of the ejection velocity on the galaxy circular velocity
also derives from the \citet{muratov15a} high-resolution simulations.
The circular velocities are determined on the fly using a fast
friends-of-friends finder \citep{dave16a}.

Feedback from longer-lived stars (e.g. asymptotic giant branch stars
[AGB] and Type 1a supernovae) are included as well (following
\citet{bruzual03a} stellar evolution tracks with a \citet{chabrier03a}
initial mass function).  We track the evolution of 11 elements: H, He,
C, N, O, Ne, Mg, Si, S, Ca and Fe.  The yields for SNe Ia are taken
from \citet{iwamoto99a}, assuming $1.4 M_\odot$ of returned mass per
supernova event.  AGB yields are drawn from the \citet{oppenheimer08a}
lookup tables.  Type II supernovae yields derive from
\citet{nomoto06a} parameterizations, though are reduced by $50\%$
owing to studies that find these yields return galaxies with
metallicities roughly a factor $\sim 2$ too large at a fixed stellar
mass when compared to the observed mass-metallicity relation
\citep{dave11a}.

The hydrodynamic simulations all use {\sc gizmo} in the meshless
finite mass mode (MFM), with a cubic spline of $64$ neighbors in the
MFM hydrodynamics.  This kernel is used to define the volume partition
between gas elements, and the faces therefore over which the
hydrodynamics is solved with the Riemann solver. Our final particle
masses are as follows.  The dark matter particle masses are $M_{\rm
  DM} = 1\times10^6 h^{-1} M_\odot$, and baryon particle masses are
$M_{\rm b} = 1.9 \times 10^{5} h^{-1} M_\odot$.  We employ adaptive
gravitational softening for all particles throughout the simulation
with minimum force softening lengths of $12, 3$ and $3$ pc for dark
matter, gas and star particles respectively.

\begin{table*}
	\centering
	\caption{Description of the simulated galaxies.  We present
          all masses at $z=2$, regardless of what the final simulation
          redshift is in order to facilitate comparison between the
          models.  These are ordered by either their actual, or
          expected $z=2$ halo mass.}
	\label{table:ics}
	\begin{tabular}{lcccc}
		\hline
		Name &  $M_{\rm *,central} ({\rm M_\odot}; z=2$) & $M_{\rm *,halo} ({\rm M_\odot}; z=2$)&$M_{\rm DM} ({\rm M_\odot}; z=2)$ & $z_{final}$\\
		\hline
                mz0     & $8.4 \times 10^{10}$ & $4.1 \times 10^{11}$&$4.1 \times 10^{13}$ & 2.15\\
                mz5     & $6.9 \times 10^{10}$ & $8.3 \times 10^{11}$&$6.3 \times 10^{13}$ & 2\\ 
                mz45    & $1.3 \times 10^{10}$ &$1.3 \times 10^{11}$& $3.7 \times 10^{13}$ & 2\\
		mz10    & $6.8 \times 10^{10}$ & $1.6 \times 10^{11}$&$1.1 \times 10^{13}$ & 2\\
                z0mz352          & $2.7 \times 10^9$ &$7.2 \times 10^{9}$& $9.2 \times 10^{11}$ & 0\\
                z0mz401         & $2.4 \times 10^9$ &$3.8 \times 10^{9}$& $5.8 \times 10^{11}$ & 0\\
                z0mz287     & $1.0 \times 10^{9}$ & $2.3 \times 10^{9}$&$2.9 \times 10^{11}$ & 0.65\\
                z0mz374         & $1.0 \times 10^{8}$& $2.9 \times 10^{8}$& $1.8 \times 10^{11}$ & 0.4\\
 		\hline
	\end{tabular}
\end{table*}

\subsection{Dust Radiative Transfer Models}
We generate the stellar SEDs and subsequent dust radiative transfer
with {\sc powderday}, a code package that wraps {\sc fsps}
\citep{conroy09b,conroy10a,conroy10b}, {\sc hyperion}
\citep{robitaille11a} and {\sc yt} \citep{turk11a}.  This process is
performed on all snapshots at redshifts $z<10$ in post-processing.

We generate the stellar SEDs using {\sc fsps}, and in particular,
their python hooks {\sc
  python-fsps}\footnote{\url{https://github.com/dfm/python-fsps}}.
The SEDs are calculated for each star particle as a simple stellar
population based on its age and metallicity, which is taken directly
from the hydrodynamic simulations.  We assume a \citet{kroupa02a}
stellar IMF for the stellar SED generation, and the Padova stellar
isochrones \citep{marigo07a,marigo08a}.  The sum of these stellar SEDs
comprise the intrinsic stellar SED for any given model galaxy
snapshot.

We then calculate the attenuation these SEDs experience by performing
dust radiative transfer calculations.  Functionally, the radiative
transfer must occur on a grid.  We therefore project the metal mass
from the hydrodynamic simulations onto a $200$ kpc octree grid
centered on the central galaxy's center of mass.  Each cell is
recursively subdivided until a maximum of $64$ particles are in a
cell.  The dust mass is assumed to be a constant $0.4\times$ the metal
mass within a cell, as motivated by observational constraints from
both local epoch galaxies and those at high-redshift
\citep{dwek98a,vladilo98a,watson11a}.

The radiative transfer from stellar sources occurs in 3 dimensions in
a Monte Carlo fashion using the dust radiative transfer code {\sc
  hyperion} \citep{robitaille11a}.  {\sc hyperion} uses the
\citet{lucy99a} algorithm for determining the converged equilibrium
dust temperature and radiation field.  Here, radiation is emitted from
stellar sources and then absorbed, scattered, and re-emitted from dust
in each cell in the octree grid.  This process is iterated upon until
the dust temperature has converged; convergence is determined when the
energy absorbed by $99$ per cent of the cells has changed by less than
$1$ per cent between iterations.  We compute the emergent intensity
for $9$ isotropic viewing angles around our model galaxies.

 For the dust grain properties themselves, we utilize the
 carbonaceous-silicate grain model from \citet{draine07a} with a size
 distribution from \citet{weingartner01a}.  This model uses the
 \citet{draine03a} renormalization relative to H and we assume $R_{\rm
   V} \equiv A_{\rm V}/E(B-V) = 3.15$.  This curve is shown as a
 reference in Figure~\ref{figure:literature}.

\section{Insight from Simplified Toy Models}
\label{section:sps_models}

\begin{figure*}
\begin{tabular}{cc}
\hspace{-0.25cm}
\includegraphics[width=\columnwidth]{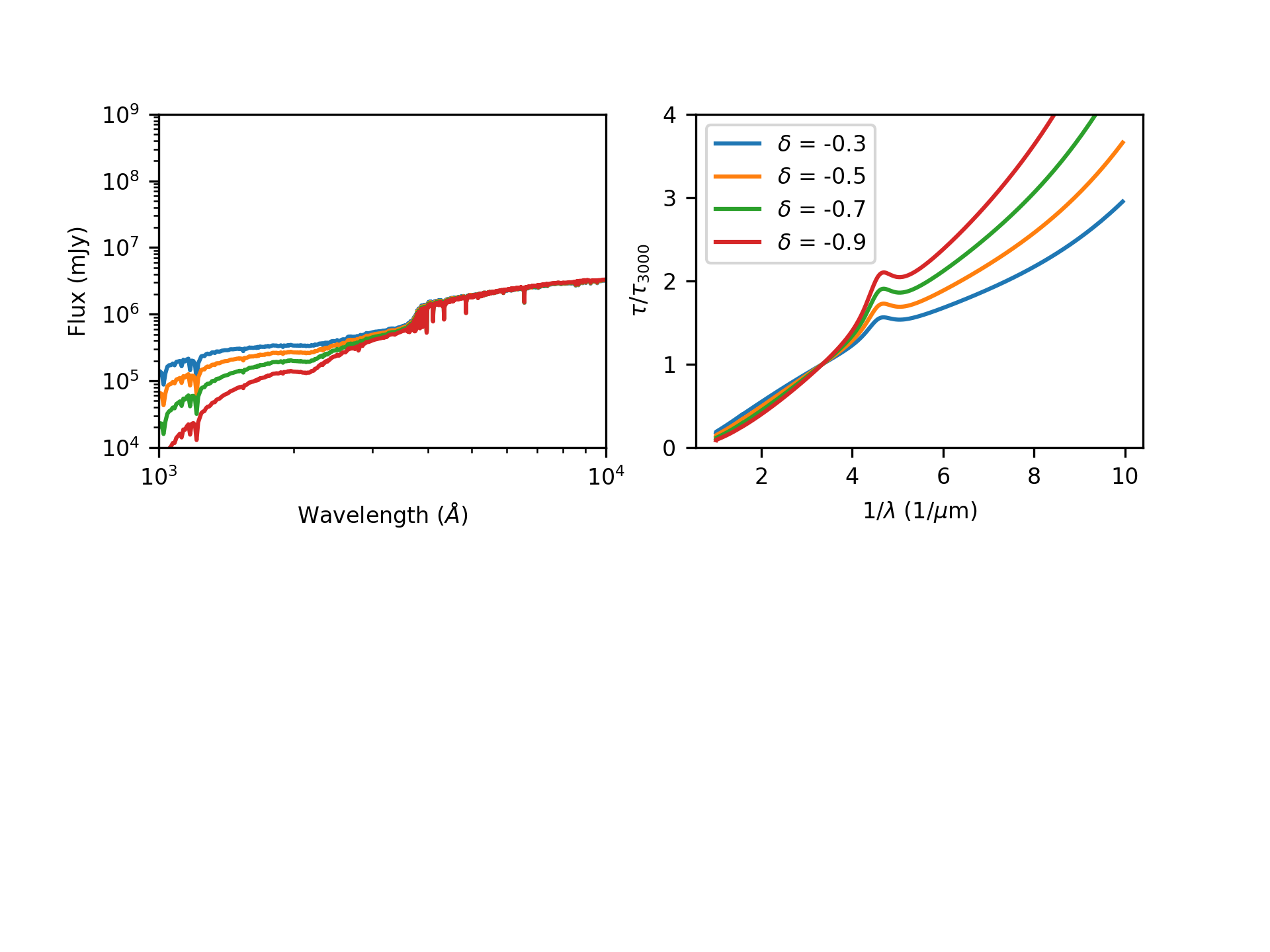}
\end{tabular}
\vspace{-6cm}
    \caption{ Model SEDs and dust attenuation curves resulting from
      {\sc fsps} stellar population synthesis calculations in which we
      vary the underlying dust extinction properties. In
      Figures~\ref{figure:delta_variations}-\ref{figure:stellar_ages_sfh},
      we intend to demonstrate that there are three fundamental drivers
      of variations in normalized attenuation laws: variations in the
      dust extinction properties, variations in the star-dust
      geometry, and stellar age effects.  Here, we concentrate on
      extinction properties.  The stellar populations shown here are
      modeled as having a constant star formation history (with SFR =
      $1 M_\odot$ yr$^{-1}$) and a stellar age of 13 Gyr.  The left
      panel shows the UV-optical SED of a stellar population hidden
      behind a dust screen with a \citet{kriek13a} extinction curve,
      and a varying powerlaw index for that curve ($\delta$).  The
      right panel shows the modeled attenuation curves with these
      varying extinction curve indices.  As is straight forward,
      variations in the dust extinction properties drive commensurate
      variations in the observed attenuation
      curves.  \label{figure:delta_variations}}
\end{figure*}

\begin{figure*}
\begin{tabular}{cc}
\hspace{-0.25cm}
\includegraphics[width=\columnwidth]{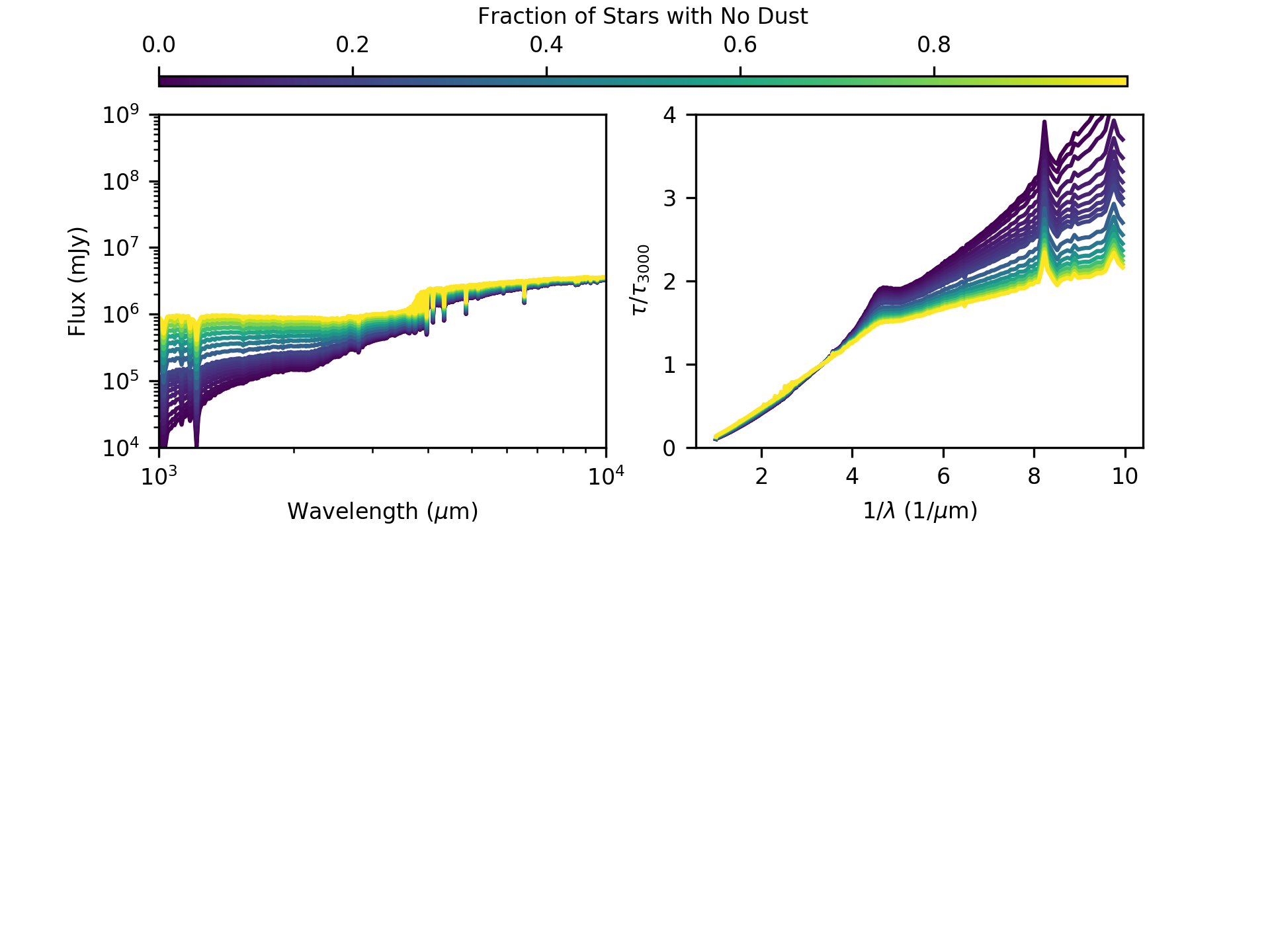}
\end{tabular}
\vspace{-6cm}
\caption{ Model SEDs and dust attenuation curves resulting from {\sc
    fsps} stellar population synthesis calculations in which we vary
  the underlying dust extinction properties. In
  Figures~\ref{figure:delta_variations}-\ref{figure:stellar_ages_sfh},
  we intend to demonstrate that there are three fundamental drivers of
  variations in normalized attenuation laws: variations in the dust
  extinction properties, variations in the star-dust geometry, and
  stellar age effects.  Here, we concentrate on the star-dust
  geometry, and model the population similarly as in
  Figure~\ref{figure:delta_variations}.  The left panel shows the
  model SEDs with a varied fraction of stars that are not attenuated
  by this dust (i.e. a fraction of $0$ corresponds to all stars
  residing behind the dust screen, and a fraction of $1$ corresponds
  to no obscuration).  The right panel shows the resultant attenuation
  curves.  As the fraction of unattenuated star light increases in
  galaxies, the attenuation curves become shallower.  This effect will
  map to increasingly complex star-dust geometries in high-redshift
  galaxies, with reduced optical depth sightlines to stars.
  \label{figure:frac_nodust}}
\end{figure*}

\begin{figure}
\begin{tabular}{cc}
\hspace{-0.25cm}
\includegraphics[width=\columnwidth]{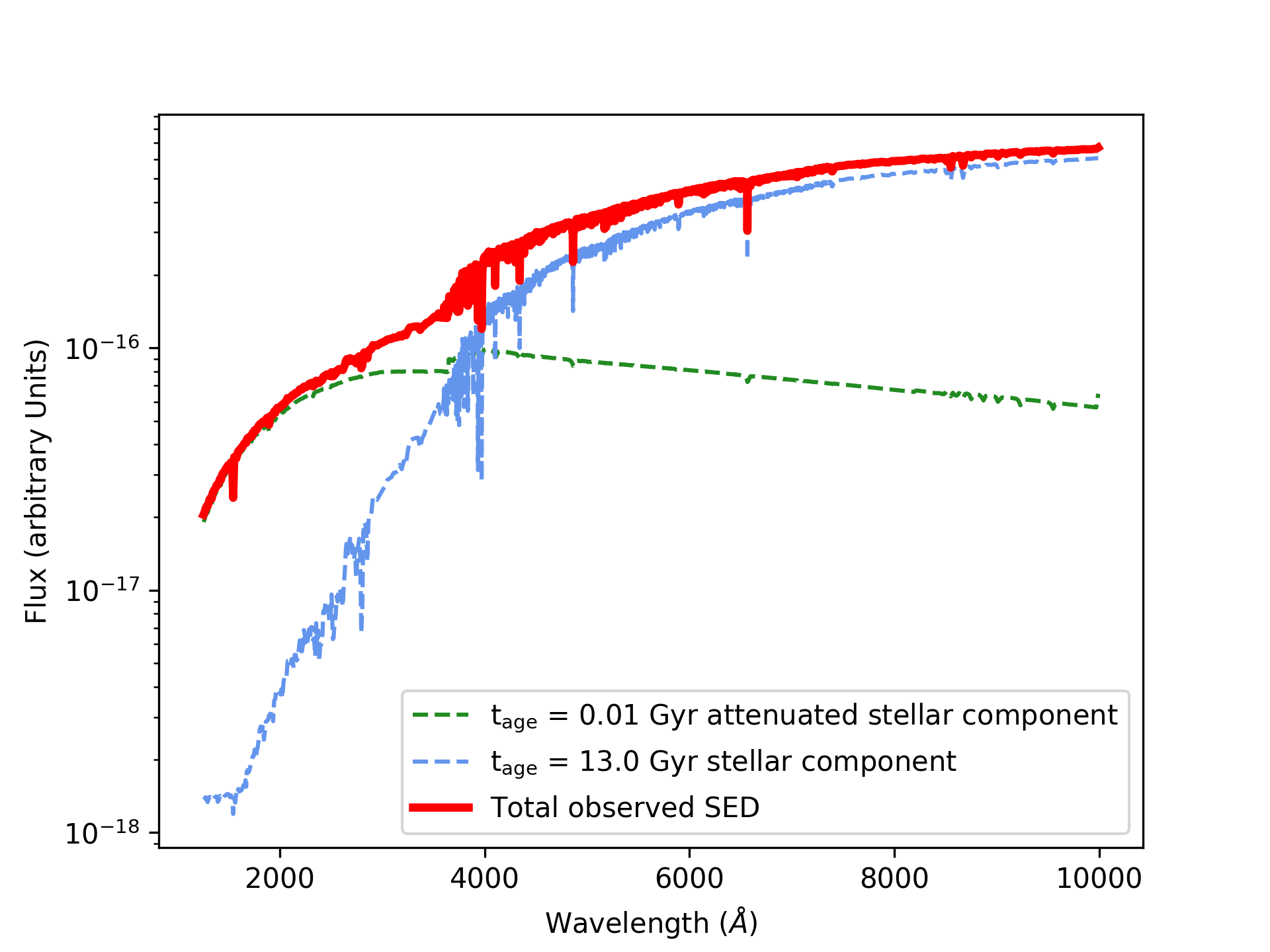}
\end{tabular}
    \caption{Results from population synthesis experiment aimed at
      simulating a galaxy dominated by old stars, with a minor amount
      of residual star formation.  In this scenario, we see that the
      UV is dominated by young stars (that live in attenuating birth
      clouds), while the older stellar population dominates the
      optical.  When comparing to
      Figure~\ref{figure:stellar_ages_sfh}, we see that this scenario
      drives very steep attenuation curves.  In detail, we create a
      composite stellar population in which $99\%$ of the stellar mass
      in this stellar population is made up of a $13$ Gyr population,
      and the remaining $1\%$ is made up of a $0.01$ Gyr population.
      The young stellar population sees a \citet{charlot00a} dust
      screen, while the older population does not.  
      \label{figure:gergo}}
\end{figure}

\begin{figure*}
\begin{tabular}{cc}
\hspace{-0.25cm}
\includegraphics[width=\columnwidth]{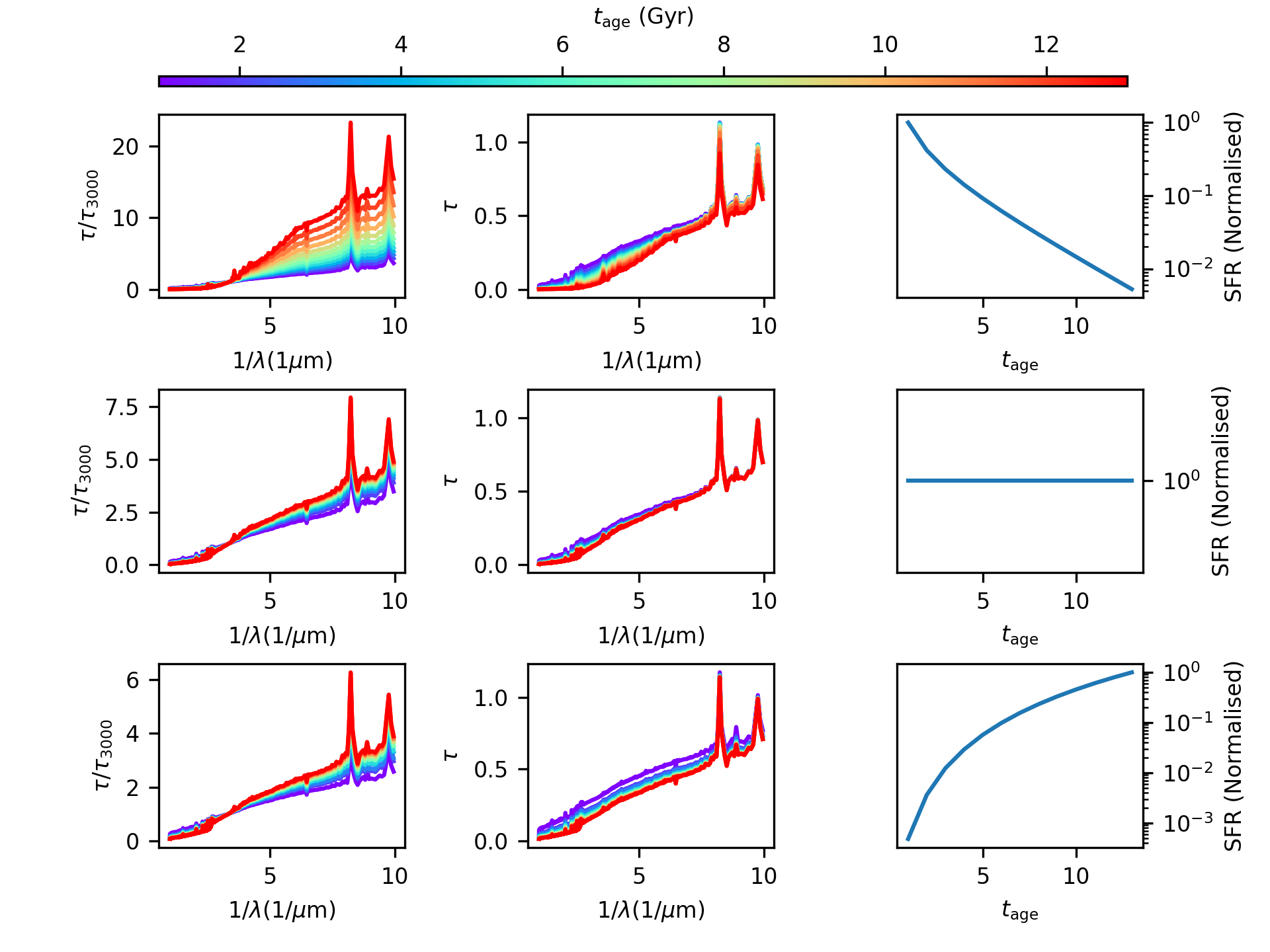}
\end{tabular}
    \caption{Results of numerical stellar population synthesis
      experiments in which we consider the impact of varying star
      formation histories and stellar ages on the derived dust
      attenuation curve.  In short: galaxies whose mass is dominated
      by older stellar populations, though have residual star
      formation have relatively steep normalized attenuation curves.
      This is because the optical light, dominated by the older
      populations, see little dust, while the UV (coming from new
      stars) are attenuated by dust associated with their birthclouds.
      We demonstrate this by creating stellar populations with a range
      of star formation histories: exponentially decreasing (top row),
      constant (middle row) and rising star formation history (bottom
      row). These star formation histories are shown explicitly in the
      last column.  The stars are placed behind a dust screen, with an
      escape time of $10$ Myr (meaning any star older than this age
      will not reside in birthclouds anymore).  The left column shows
      the attenuation curves normalized at $3000 \angstrom$, and the second
      column shows the absolute attenuation curve (i.e. not normalized
     at $3000 \angstrom$).  The third column shows the UV-near infrared
      SEDs for both the youngest and oldest stellar age in the SFH.
      Owing to a combination of the mixing of young and old stars with
      the ISM and the relative UV and optical flux emitted by young
      and old stars, the stellar age of a galaxy can serve as a proxy
      for the underlying effects that drive variations in attenuation
      law slopes.  See text for details.
      \label{figure:stellar_ages_sfh}}
\end{figure*}

We begin the analysis for this paper by first conducting a series of
simplified stellar population synthesis numerical experiments.  It is
from these that we will build the physical intuition necessary to understand
the results of significantly more complex systems (i.e. the
cosmological zoom galaxy formation simulations).

The relevant parameters in the dust attenuation curve for SED modeling
are its normalization and its shape.  The overall normalization is
trivially a function of the total dust column density seen by stars.
As a result, we do not explore this further, and concentrate hereafter
on the shape of the attenuation curve.

When considering the shape (i.e. slope) of the attenuation curve, we
assert that three principle physical drivers are at play: (i) the
shape of the underlying extinction curve (which, itself, is dictated
by the dust grain size distribution), the (ii) the star-dust geometry,
and the (iii) the average stellar age.  We demonstrate the first
(grain size variations) effect in
Figure~\ref{figure:delta_variations}, the second effect (star-dust
geometry) in Figure~\ref{figure:frac_nodust}, and the final effect
(stellar ages) in 
Figures~\ref{figure:gergo}-\ref{figure:stellar_ages_sfh}.

 In Figures~\ref{figure:delta_variations} and
 \ref{figure:frac_nodust}, we create (as an extreme example) a stellar
 population with {\sc fsps} \ \citep{conroy09b,conroy10a,conroy10b}
 with age $13$ Gyr formed with a constant $1 M_\odot$ yr$^{-1}$ star
 formation history.  Stars reside behind a dust screen whose opacity
 varies as a function of stellar age.  Following \citet{charlot00a},
 we take: \renewcommand{\arraystretch}{2}
\begin{equation}
\tau = \left\{ \begin{array}{cl} \tau_1\left(\lambda/5500 \angstrom\right)^{\delta_{\rm CF}} & t\leq 10^7 {\rm yr} \\
  \tau_2\left(\lambda/5500 \angstrom\right)^{\delta_{\rm CF}}&t > 10^7 {\rm yr}\\
\end{array}\right.\,\,,
\end{equation}
such that the dust screen in front of the stellar population in this
experiment has a variable age-dependent normalization.  Here, we set
the normalization for young stars ($t_{\rm age} < 10^7$ yr) as
$\tau_1=1$, while older stellar populations see a normalization $\tau_2 =
0.3$ \citep[see][]{conroy10a}. Following \citet{charlot00a}, we set
$\delta_{\rm CF} = 0.7$.  The strong attenuation for young stars is
intended to emulate the attenuation of starlight from newly formed
stars by their birthclouds, while the more modest attenuation for
older populations serves as a proxy for diffuse cirrus dust in
galaxies.  The extinction curve follows the 
\citet{kriek13a} derived curve\footnote{Formally, the \citet{kriek13a} curve is
    an attenuation curve.  For these models, we use it as an {\it
      extinction} curve, though note that our results are robust
    against using a wide range of extinction curves.  One aspect of
    the \citet{kriek13a} curve that is visible in
    Figure~\ref{figure:delta_variations} (though unimportant for this
    particular experiment) is that the bump strength manifestly varies
    with the slope powerlaw index.  We further note that the choice of a \citet{kriek13a} curve is {\it only} used for the numerical experiments in this section}.  

In Figure~\ref{figure:delta_variations}, we vary the slope of the
imposed \citet{kriek13a} extinction curve as a proxy for variations in
the underlying dust grain properties, and hold the dust covering
fraction fixed at unity (meaning all stars see the dust screen).  It
is evident that variations in the intrinsic extinction curve propagate
to a wide range of observed attenuation laws.  At the same time,
absent a model for the physical evolution of dust grain properties in
cosmologically evolving galaxies
\citep[e.g.][]{mckinnon16a,popping17b,hou17a}, we are forced to assume
a fixed underlying extinction curve, and do not consider variations
that originate in the physics of dust grain property variations
further.  We instead focus on the impact of the star-dust geometry and
stellar ages on the shape of the normalized attenuation curve.

In Figure~\ref{figure:frac_nodust}, we show a
similar model as in Figure~\ref{figure:delta_variations}, though fix the slope of the \citet{kriek13a} extinction
curve ($\delta = -0.7$) and instead vary the fraction of stars that do
not see any dust.  This model is intended to serve as a proxy for
increasingly complex star-dust geometries in galaxies.  A simplified
geometry where all stars are enshrouded by dust is represented by a
fraction of $0.0$, while a model in which the stellar light and dust
clouds are wholly decoupled is represented by a fraction $1.0$.  As
the star-dust geometry becomes increasingly complex, and more of the
star light is decoupled from dust, the effective attenuation curve
becomes flatter (greyer), a well known result dating back to seminal
works by \citet{witt96a}.

Note that a 'complex star-dust geometry' can encapsulate a broad range
of effects.  For example, a system that has a significant young stellar
population that dominates the UV and is enshrouded in dust, but also
a significant unobscured old star population that dominates the
optical will have a maximally steep attenuation curve.  On the other
hand, a galaxy with a significant unobscured young star population
\citep[e.g.][]{casey14b,geach16a,narayanan18a}, but obscured old star
population would have an extremely shallow (grey) attenuation curve.
Intermediate cases naturally fall between these two limits.

Because of this, alongside the typical star-dust geometry, the median
stellar age of a galaxy can impact observed dust attenuation curves as
well.  In short: galaxies with a young median stellar age have
  both their UV and optical light dominated by young stars.  Hence,
  the shape of the attenuation curve in this situation is dictated
  primarily by the fraction of obscured young stars.  Galaxies with an
  older median stellar age, however, represents a different situation.
  Here, the optical light is typically dominated by older stars, which
  tend to be free from obscuring dust, while the UV light is still
  dominated by young stars.  In this case, the ultraviolet regime in
  the attenuation curve is dictated by the fraction of obscured young
  stars, while the optical is determined by the fraction of obscured
  old stars.

 To see this explicitly, we develop a simplified population synthesis
 experiment in Figure~\ref{figure:gergo}.  Here, we have constructed a
 composite stellar population (red line) comprised of a $0.01$ Gyr
 population (green line) and a $13$ Gyr population (blue line).  The
 $0.01$ Gyr population constitutes $1$\% of the total mass, and the
 $13$ Gyr population constitutes the remaining $99\%$.  The intention
 behind this contrived experiment is to mimic a galaxy comprised
 principally (to the $99\%$ level) of old stars, with only residual
 ($1\%$ of the total mass) current star formation.  As usual in our
 population synthesis numerical experiments, the young stars are
 hidden behind a \citet{charlot00a} screen of dust with the same
 parameters used thus far.  The old stars do not suffer any dust
 attenuation.

Evident in Figure~\ref{figure:gergo} is that the new stars (that
suffer attenuation from a dust screen) dominate the SED at FUV
wavelengths, while the older stellar population dominates the
optical. Because the older stars do not see a dust screen, this will
result in increased transparency at optical wavelengths, though
depressed emission at FUV wavelengths (thanks to the young stars
hidden behind the \citet{charlot00a} screen).  This scenario can
therefore play an important role in determining the shape of
normalized attenuation curves.


To demonstrate the impact of stellar ages on the normalized
  attenuation curves even further, in
  Figure~\ref{figure:stellar_ages_sfh}, we increase the complexity of
  the numerical experiments developed in Figure~\ref{figure:gergo},
  and model three different star formation histories in population
  synthesis models.  The top row shows an exponentially declining
SFH, the middle row a constant SFH, and the bottom row a rising SFH;
the SFH for each is quantified in the last (third) column.  As in all
simplified population synthesis experiments developed thus far in this
paper, in Figure~\ref{figure:stellar_ages_sfh}, we utilize the
\citet{charlot00a} type dust model, where young stars reside behind a
dust screen, though evolved stars do not.

In the first and second columns, we show the normalized and absolute
attenuation curves for a range of stellar ages.  The third column
shows the star formation history.  Let us first consider the case of
the exponentially declining star formation history (row 1) as an
instructive example.  When examining the first column of the first row
(normalized attenuation curves), we see that the models with the
oldest ages have the steepest normalized curves.  The reason for this
becomes evident in the second column (absolute attenuation curve).  As
the star formation history evolves in the top row of
Figure~\ref{figure:stellar_ages_sfh}, the obscuration of the sources
that dominate the far ultraviolet (FUV; $x\ga6$) photons doesn't
change (thanks to the assumption of \citet{charlot00a} birth clouds),
but that of the optical photons does.  Galaxies with more evolved
stellar populations still have their UV fluxes dominated by young
stars, though their optical fluxes derive instead from older stellar
populations.  What this means is, if young stars are generally more
obscured than older stars (in the stellar population synthesis models
explored here, this is manifestly enforced), then galaxies with young
median stellar ages will see significant obscuration in both the UV,
as well as the optical.  Galaxies that are more evolved will still see
significant obscuration in the UV (as the young stars are still
enshrouded in dust), though the older stars that dominate the optical
have lower optical depths (column 2 of
Figure~\ref{figure:stellar_ages_sfh}), and therefore steeper
normalized attenuation curves.  

The same effects are visible in the second and third rows of
Figure~\ref{figure:stellar_ages_sfh} (constant and rising SFH,
respectively).  This said, the dispersion in attenuation curve slopes
is more modest compared to the exponentially declining history owing
to the mixed contribution to optical light by both young and older
stellar populations in the most evolved ($t_{\rm age} = 13 $ Gyr)
stellar age bin.



\section{Dust Attenuation Curves in Cosmological Galaxy Formation Simulations}
\label{section:results}
With the insight we have built from our simplified stellar population
synthesis models in \S~\ref{section:sps_models}, we now turn to the
modeled curves in our cosmological zoom galaxy formation simulations.
As a reminder: going forward we will hold the underlying dust
extinction properties fixed with a \citet{weingartner01a} size
distribution.  Beyond this, because the inclusion of
  subresolution birth clouds depends on tunable free-parameters that
  can impact the amount of attenuated light
  \citep[e.g.][]{narayanan09a,narayanan10a,hayward13a} we will
abandon the usage of any subresolution 'birthcloud' model in the zoom
simulations.   All attenuation seen by stars will be on larger scales
($\ga 10$ pc) from the geometry of the galaxy itself.

\subsection{Diversity in Dust Attenuation Curves}
\begin{figure}
\begin{tabular}{cc}
\hspace{-0.25cm}
\includegraphics[width=\columnwidth]{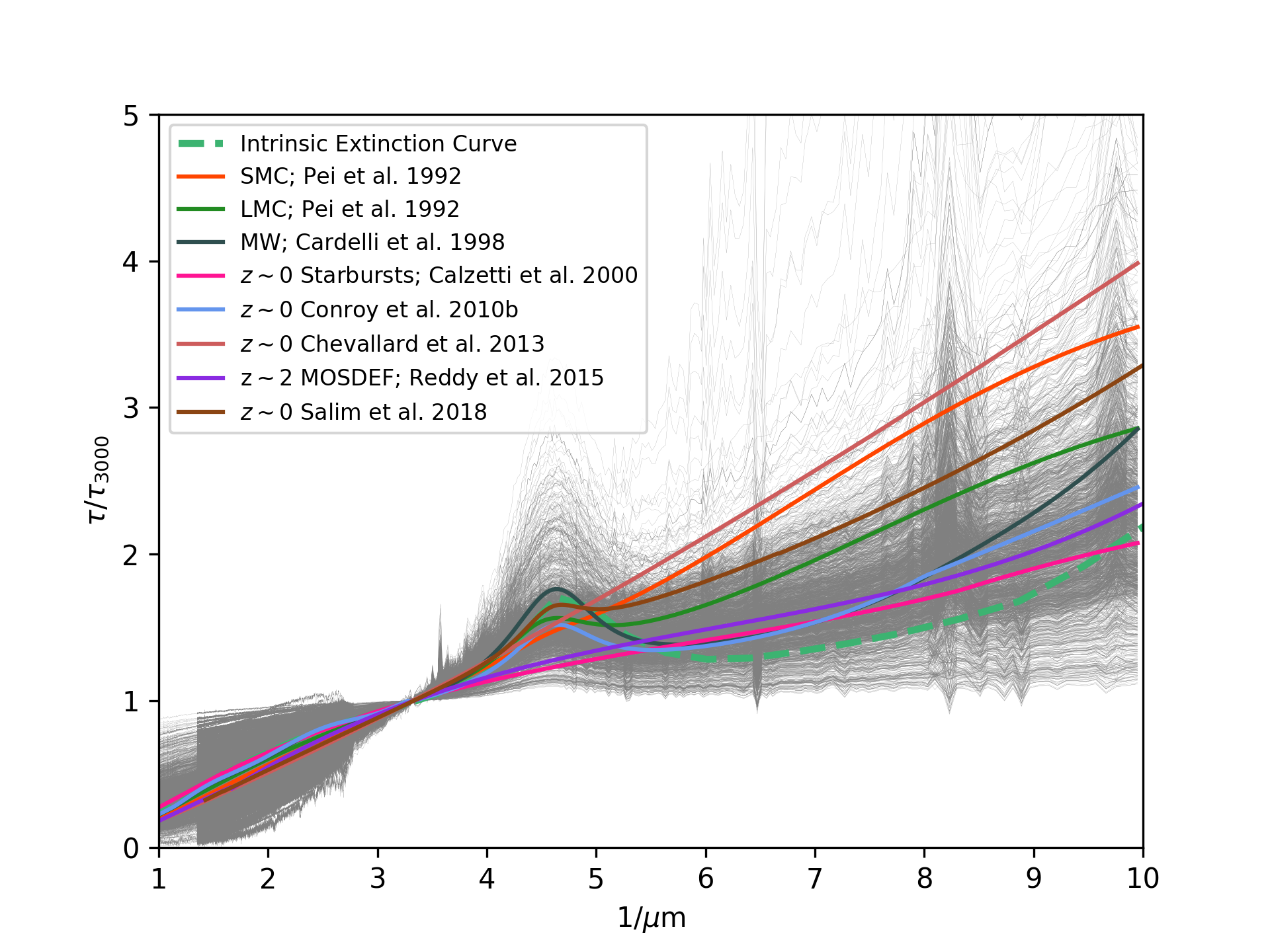}
\end{tabular}
\caption{Diversity of attenuation laws for the zoom galaxy formation
  simulations examined in this paper.  The curves are normalized by
  their $3000 \angstrom$ optical depth.  As is evident, a wide
  diversity of slopes and bump strengths are evident, with curves that
  are both steeper and greyer than standard literature assumptions.
  We show our intrinsic extinction curve via the green dashed line.
  Even with a constant extinction curve in all of our simulations, a
  diverse range of attenuation curves can
  emerge.\label{figure:all_curves}}
\end{figure}

To set the stage, in Figure~\ref{figure:all_curves}, we show the
diversity of attenuation curves for all snapshots of all model zoom galaxy formation
simulations examined in this paper.  The attenuation curves are
normalised by their $3000 \ \angstrom$ optical depth (which,
  hereafter, we refer to as $\tau_{3000}$).  We additionally show a
number of observationally-constrained literature attenuation curves
for reference. It is clear that a wide range of attenuation curves
emerge from these simulations, with curves both steeper and grayer
(shallower) than the standard literature assumptions.  The goal of the
remainder of this section is to unpack these curves further, and
examine in detail the physical drivers behind this diversity in
attenuation curves.

\subsection{Fitting Parameterizations}
\label{section:fitting}
In our analysis of the physical drivers of the shapes of attenuation
curves, we begin by describing formulae that we employ to fit our
model attenuation curves.  It is useful to go through this exercise
at this point, as this will allow us to introduce variables that we
can use to characterize various properties of the modeled attenuation
curve (e.g. their $2175 \angstrom$ UV bump strengths, or the
normalized slope of the attenuation curve).

Broadly, we follow a modified version of the parameterizations
described in \citet{conroy10a}, which themselves are a modified
version of the \citet{cardelli89a} fits to the average Milky Way
extinction curve.  For the $2175 \angstrom$ UV bump, we utilize a
Drude profile, and in the near ultraviolet (NUV), we follow
\citet{noll07a}.

In more detail, we first define $x$ as the inverse wavelength with
units of $1/\mu$m.  We fit the infrared ($0.3 \leq x < 2.5$)   with:
\begin{eqnarray}
  \label{equation:fit_ir}
  a(x)_{\rm IR} = a_1 \times x^{\gamma_{\rm IR}}\\
  b(x)_{\rm IR} = b_1 \times x^{\gamma_{\rm IR}}\\
  f_{\rm IR} = \left(a(x)_{\rm IR}+b(x)_{\rm IR}\right)/R_{\rm V}
\end{eqnarray}
In the optical ($2.9<x\leq2.5$), we define:
\begin{equation}
y = x-1.82
\end{equation}
and then parameterize the optical via:
\begin{eqnarray}
  a(x)_{\rm opt} = 1 + a_1y - a_2y^2 - a_3y^3 + a_4y^4 + a_5y^5 - a_6y^6 + a_7y^7\\
  b(x)_{\rm opt} = b_1y + b_2y^2 + b_3y^3 - b_4y^4 - b_5y^5 + b_6y^6-b_7y^7\\
  f_{\rm opt} = \left(a(x)_{\rm opt}+b(x)_{\rm opt}\right)/R_{\rm V}
\end{eqnarray}
The near ultraviolet (NUV) and $2175 \angstrom$ UV bump are modeled
via a \citet{calzetti00a} law, with a Lorentzian-like
Drude Profile.  The latter is given by:
\begin{equation}
  \label{equation:drude}
D_{\lambda_0,\gamma,E_{\rm bump}} = \frac{E_{\rm bump}\lambda^2\gamma^2}{\left(\lambda^2-\lambda_0^2\right)^2 + \lambda^2\gamma^2}
\end{equation}
where $\lambda_0 = 2175 \angstrom$ is the central wavelength of the
bump, $\gamma$ is the width, and $E_{\rm bump}$ is the amplitude
\citep{fitzpatrick07a,noll09a}.  Formally, the fit for the NUV and
bump take the form:
\begin{equation}
\label{equation:nuv}
  f_{\rm UV,bump} = A_{\rm  NUV \ norm}\left[(k(\lambda)+D_{\lambda_0,\gamma,{\rm E_{\rm bump}}}) \frac{1-1.12 c_{\rm R}}{R_{\rm V}}+1 \right]\left(\frac{\lambda}{5500 \angstrom}\right)^{\delta_{\rm NUV}}
\end{equation}
Here, $A_{\rm FUV \ norm}$ is a normalization of the fit, and $k(\lambda)$
is the \citet{calzetti00a} law over the wavelengths of interest.  The
$(\lambda/5500)^\delta_{\rm NUV}$ term allows for an arbitrary tilt to the curve
in this wavelength regime, and the $1-1.12c_R/R_V$ term compensates
for how $R_{\rm V}$ will change in the original Calzetti law due to
the introduction of a tilt \citet{noll09a}.

Finally, we model the far ultraviolet (FUV) between $5.9\leq x<10$ via:
\begin{eqnarray}
  \label{equation:fit_fuv}
  f_a = a_{\rm 1,FUV}\left(x-5.9\right)^2 - a_{\rm 2,FUV}\left(x-5.9\right)^3\\
  f_b = b_{\rm 1,FUV}\left(x-5.9\right)^2 + b_{\rm 2,FUV}\left(x-5.9\right)^3\\
  a(x) = 1.752 - 0.316\times x - \frac{0.014 B}{\left(x-4.67\right)^2+0.341}+f_a\\
  b(x) = -3.09 + 1.825\times x + \frac{1.206B}{\left(x-4.62\right)^2 + 0.263}+f_b\\
  f_{\rm FUV} = \left(\frac{a(x)+b(x)}{R_{\rm V}}\right)A_{\rm FUV,norm}
\end{eqnarray}
In other parameterizations of attenuation curves
\citep[e.g.][]{conroy10a}, $B$ is a parameter describing the bump
strength.  Because we employ the \citet{noll09a} model for the NUV and
bump, here (in the FUV), $B$ serves simply as another free parameter
in the fit.  In this wavelength regime, we linearly interpolate over
the Ly$\alpha$ line to avoid complications during fitting.

\begin{figure}
  \begin{tabular}{cc}
    \hspace{-0.25cm}
    \includegraphics[width=\columnwidth]{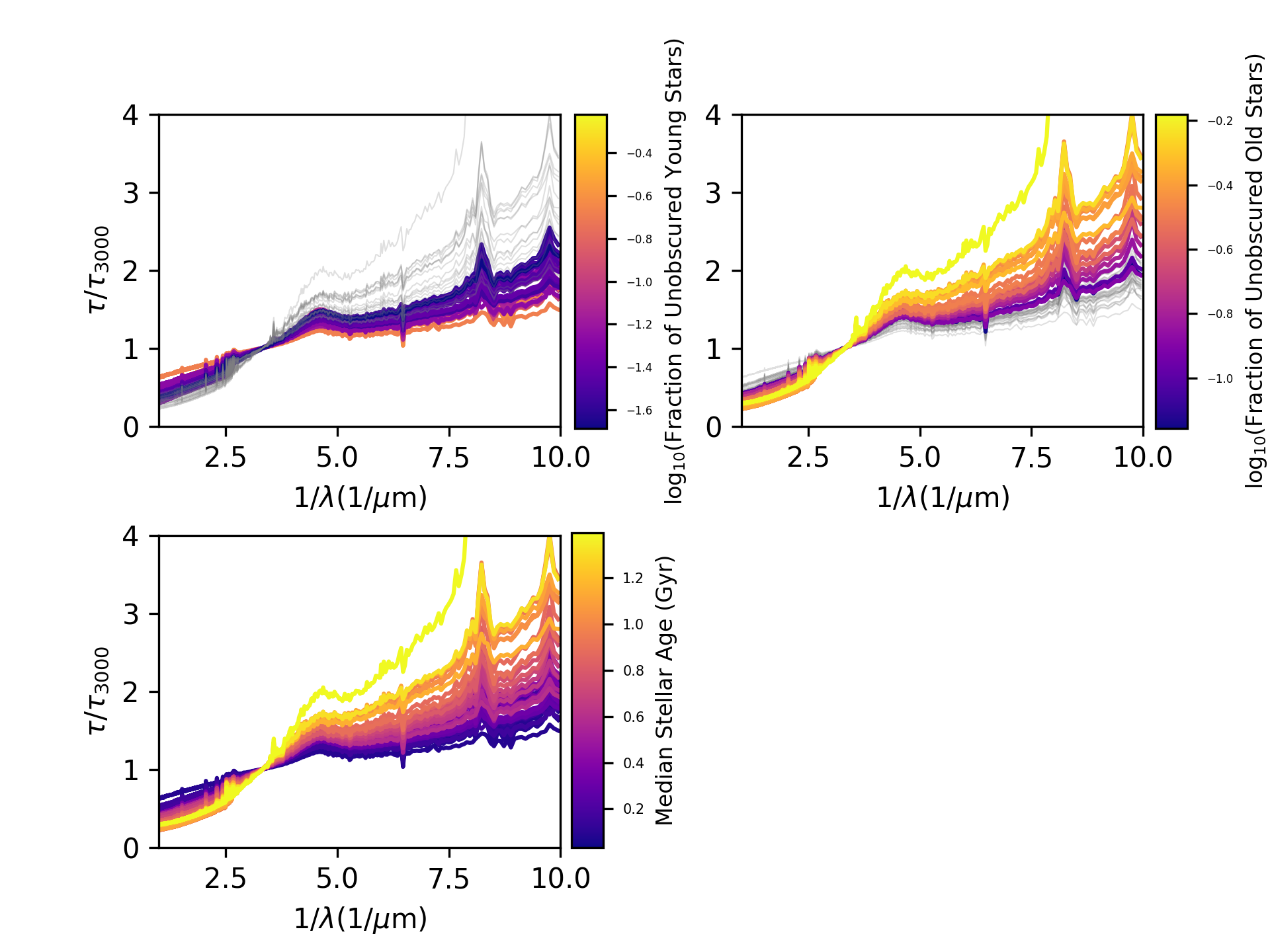}
  \end{tabular}
  \caption{Dust attenuation curves for a representative galaxy zoom
    model (mz10).  These are color-coded by the fraction of young
    stars unobscured by dust (top left), the fraction of old stars
    unobscured by dust (top right), and the median stellar age (bottom
    left).  {\bf Top Left:} Galaxies with more obscuration of young
    stars tend to have steeper attenuation curves, while those with a
    more complex geometry (and a larger fraction of unobscured young
    stars) have flatter (greyer) attenuation curves.  To minimize
    contamination by older stellar populations, we only color-code
    galaxies with $t_{\rm age} < 0.5 $ Gyr, though show all snapshots
    in grey.  {\bf Top Right:} For galaxies in which old stars
    dominate the optical luminosity, larger fractions of unobscured
    old stars result in steeper attenuation curves.  To minimize
    contamination by younger stellar populations, we only color-code
    galaxies with $t_{\rm age} > 0.5 $ Gyr, though show all snapshots
    in grey.  {\bf Bottom Left:} Akin to the top right, older stellar
    populations have the majority of their optical flux emitted from
    older stars that are relatively decoupled from dust (while the UV
    emission is still emitted from young stars more co-spatial with
    dust).  There is therefore an increased $\tau_{\rm UV}/\tau_{\rm
      V}$ compared to a situation where the UV and optical are both
    dominated by young stars.  Consequently, galaxies with older
    stellar populations exhibit steeper attenuation curves.
    \label{figure:tilt_multiplot}}

\end{figure}

We note that while a piece-wise fit to attenuation curves akin
  to what is presented here has been used regularly in the literature
  \citep[e.g.][]{cardelli89a,conroy10a}, a number of studies in the
  literature make use of a modified \citet{calzetti00a} curve over the
  entire $x=1-10/\mu$m wavelength range:
\begin{equation}
  \label{equation:modified_calzetti}
A_{\lambda} = \Gamma \left(k'_{\rm cal} + D\left(\lambda\right)\right) +\left(\frac{\lambda}{\lambda_{\rm V}}\right)^{\delta_{\rm cal}}
\end{equation}
where $k'_{\rm cal}$ is the \citet{calzetti00a} relation,
  $\delta_{\rm cal}$ is the index for a power-law modification to this
  relation, $D\left(\lambda\right)$ is the normal Drude profile (to
  represent the $2175 \angstrom$ bump), and $\Gamma$ is a constant
  free parameter.  Various forms of this relation exist in the
  literature, where the constant may be multiplied by all or only some
  of the terms in Equation~\ref{equation:modified_calzetti}.  $\Gamma$
  is a constant that typically relates either $R_{\rm V}$ or $A_{\rm
    V}$ to the ratio of total to selective extinction for the
  \citet{calzetti00a} relation (i.e. $R_{\rm V,cal} = 4.05$), but the
  exact implementation of this also varies in the literature
  \citep[e.g.][]{noll09a,kriek13a,salmon16a,salim18a}.  Similarly, not
  all authors include the Drude representation of a bump. In order to
  compare against observations, we will find it useful at times in
  this paper to employ Equation~\ref{equation:modified_calzetti} to
  fit over the entire wavelength regime of the attenuation curve.  In
  this case, we will note the power-law indices used in this case as
  $\delta_{\rm cal}$, instead of $\delta_{\rm NUV}$, which we shall
  reserve for our normal fitting procedure.

Beyond this, some authors fix the width ($\gamma$) of the Drude
  profile representing the bump strength.  In this situation, we denote the normalization of the bump as $E_{\rm bump,KC}$, and the width of the bump as $\gamma_{\rm const}$:
\begin{equation}
  \label{equation:ebumpkc}
D_{\lambda_0,\gamma,E_{\rm bump}} = \frac{E_{\rm
    bump,KC}\lambda^2\gamma_{\rm
    const}^2}{\left(\lambda^2-\lambda_0^2\right)^2 + (\lambda^2 *
  \left(\gamma_{\rm const}\right)^2)}
\end{equation}
For clarity, we collect the fitting variables
  from this section that we will employ to compare with observations
  in Table~\ref{table:fit}.

\begin{table*}
	\centering
	\caption{Definitions of fitting variables that will be used to characterize the shapes of attenuation curves in this paper.}
	\label{table:fit}
	\begin{tabular}{ccc}
		\hline
		Fitting Variable &  Definition & Equation\\
		\hline
                $\delta_{\rm NUV}$     &  Power law index in NUV wavelength range & Equation~\ref{equation:nuv} \\
                $\delta_{\rm cal}$     & Power law index of modified \citet{calzetti00a} curve & Equation~\ref{equation:modified_calzetti} \\
                $\int_{\rm NUV} D_{\lambda_0,\gamma,E_{\rm bump}}$ & Integral of Drude Profile in NUV: bump strength & Equation~\ref{equation:drude}\\
                $E_{\rm bump}$ & Normalization of UV bump, given a variable bump width $\gamma$ & Equation~\ref{equation:drude}\\
                $E_{\rm bump,KC}$ & Normalization of UV bump, assuming a constant bump width, $\gamma_{\rm const}$& Equation~\ref{equation:ebumpkc}\\
                $\gamma_{\rm const}$ & Value of constant bump width for $E_{\rm bump,KC}$ & Equation~\ref{equation:ebumpkc}\\
 		\hline
	\end{tabular}
\end{table*}

\subsection{What Sets the Slope of Dust Attenuation Curves?}
\label{section:slope}
Building from our intuition in \S~\ref{section:sps_models}, we examine
the physical drivers of the slopes of our model dust attenuation
curves.  Fundamentally, the principal driver of variations in the
slope of attenuation curves is the star-dust geometry.  Recalling
\S~\ref{section:sps_models}, the steepest attenuation curves arise
from galaxies with a significant fraction of young stars (dominating
the emitted UV flux) obscured by dust, but old stars (dominating the
optical flux) that are not obscured.  A more mixed geometry (i.e. more
old stars obscured by dust, or more young stars decoupled from dust)
will flatten the attenuation curve from this extreme limit.  In
\S~\ref{section:sps_models}, we showed this explicitly with simplified
population synthesis models; we now explore these effects in bona fide
cosmological galaxy formation simulations.  We center this discussion
around Figure~\ref{figure:tilt_multiplot} where we show various
incarnations of the attenuation curves of an example model galaxy,
model mz10.

In the top left panel of Figure~\ref{figure:tilt_multiplot}, we show
the normalized attenuation curves for model mz10, color-coded by
their fraction of unobscured young stars.  This quantity is determined
by calculating the fraction of stellar mass in the form of young
($t_{\rm age} < 50 $ Myr) stars that have no dust within $250$ pc.  In
other words, larger values of this fraction correspond with
increasingly decoupled geometry between young stars and dust in
galaxies.  To minimize the complicating effects of older stellar
populations (an effect we will return to shortly), we only plot the
attenuation curves for galaxies with a median stellar age (by mass) $<0.5$ Gyr;
this said, in light grey we show the attenuation curve for all
snapshots of this model (i.e. of all median stellar ages).  While the dynamic
range in attenuation curve slopes is relatively small, it is clear
from the top left panel of Figure~\ref{figure:tilt_multiplot} that a larger fraction of
unobscured young stars results in a flatter dust attenuation curve.

Similarly, the geometry between old stars and dust also plays a role
in the tilt of observed attenuation curves.  To demonstrate this, in
the top right panel of Figure~\ref{figure:tilt_multiplot}, we show the
same attenuation curves as shown in the top left panel, though in this
case we color-code the attenuation curves by their fraction of
unobscured old stars.  As in the top left panel, we define
'unobscured' as having no dust within $250$ pc, and old stars as those
with $t_{\rm age} > 50$ Myr.  To minimize contamination by young
stellar populations, we only include galaxies with median stellar age
$> 0.5$ Gyr, though show the curves for all snapshots in light grey.
As in the simplified stellar population synthesis models presented in
Figure~\ref{figure:stellar_ages_sfh}, galaxies that contain a
significant amount of old stars that are unobscured by dust have steep
attenuation curves, while those that have larger obscuration of old
stars have flatter (greyer) attenuation curves.

Why are the curves in the top left panel of
Figure~\ref{figure:tilt_multiplot} that are associated with young
galaxies uniformly shallower (greyer) than the older stellar age
curves on the top right of Figure~\ref{figure:tilt_multiplot}?  As
shown in Figures~\ref{figure:gergo}-\ref{figure:stellar_ages_sfh}, in galaxies where
the median stellar age is relatively young, the UV and optical light
are both dominated by newly formed stars.  If the light from these
stars is attenuated by dust, then {\it both} the optical and UV
emission from the galaxy will be attenuated.  As a result, these
curves will not be as steep as a situation where young stars (that are
obscured) dominate the UV emission, but old stars are relatively
unobscured (a situation described by the top right panel of
Figure~\ref{figure:stellar_ages_sfh}.

What this results in is a situation where galaxies with older stellar
ages, on average, have steeper attenuation curves.  In the bottom left
panel of Figure~\ref{figure:tilt_multiplot}, we show this by plotting
the same attenuation curves as in the other two panels, though this
time color-coded by the median stellar age.  As the galaxy age
increases, so does the slope.  This is due to an increasingly large
fraction of the optical emission coming from older stars as the galaxy
ages.  Older star particles, on average, are less likely to be
associated with dust than young stars.  This same effect was
  noted by \citet{charlot00a}, who noted steeper attenuation curves
  with increasing galaxy age.

\subsection{Variations in the $2175 \ \angstrom$ Bump}
\label{section:bump}
As discussed in \S~\ref{section:introduction}, while one of the
strongest features in the attenuation curve of the Milky Way is the
bump feature observed at $\sim 2175 \angstrom$, a number of
observations show dramatic variations in bump strength in different
galaxies
\citep[e.g.][]{gordon99a,gordon00a,calzetti00a,motta02a,burgarella05a,york06a,stratta07a,noll07a,eliasdottir09a,conroy10d,buat11a,wild11a,kriek13a,scoville15a,battisti16a,salim18a}.
In this section, we utilize the model that we have developed thus far
to understand the origin of variations in the UV bump strength in
galaxy attenuation curves.

We define the strength of the UV $2175 \angstrom$ bump as the integral
over the best fit Drude profile in the NUV
(c.f. Equation~\ref{equation:drude}):
\begin{equation}
\text{Bump Strength} \equiv \int_{\rm NUV} D_{\lambda_0,\gamma,E_{\rm bump}}
\end{equation}
Note that literature definitions of the bump strength vary, with some
definitions characterizing the strength as $E_{\rm bump}$ (i.e. one
numerator term in Equation~\ref{equation:drude}), as well as
  $E_{\rm bump,KC}$, where the width of the bump is held fixed;
  c.f. Table~\ref{table:fit}.

Because the UV bump is an absorption feature, for a fixed underlying
extinction curve, reduced bump strength in the observed attenuation
curve signifies extra radiation filling in the bump.  In principle,
there are two possible sources of this extra radiation: light being
scattered into the line of sight, and unobscured sources of UV
radiation (i.e. unobscured young stars).

\begin{figure}
  \begin{tabular}{cc}
    \hspace{-0.25cm}
    \includegraphics[width=\columnwidth]{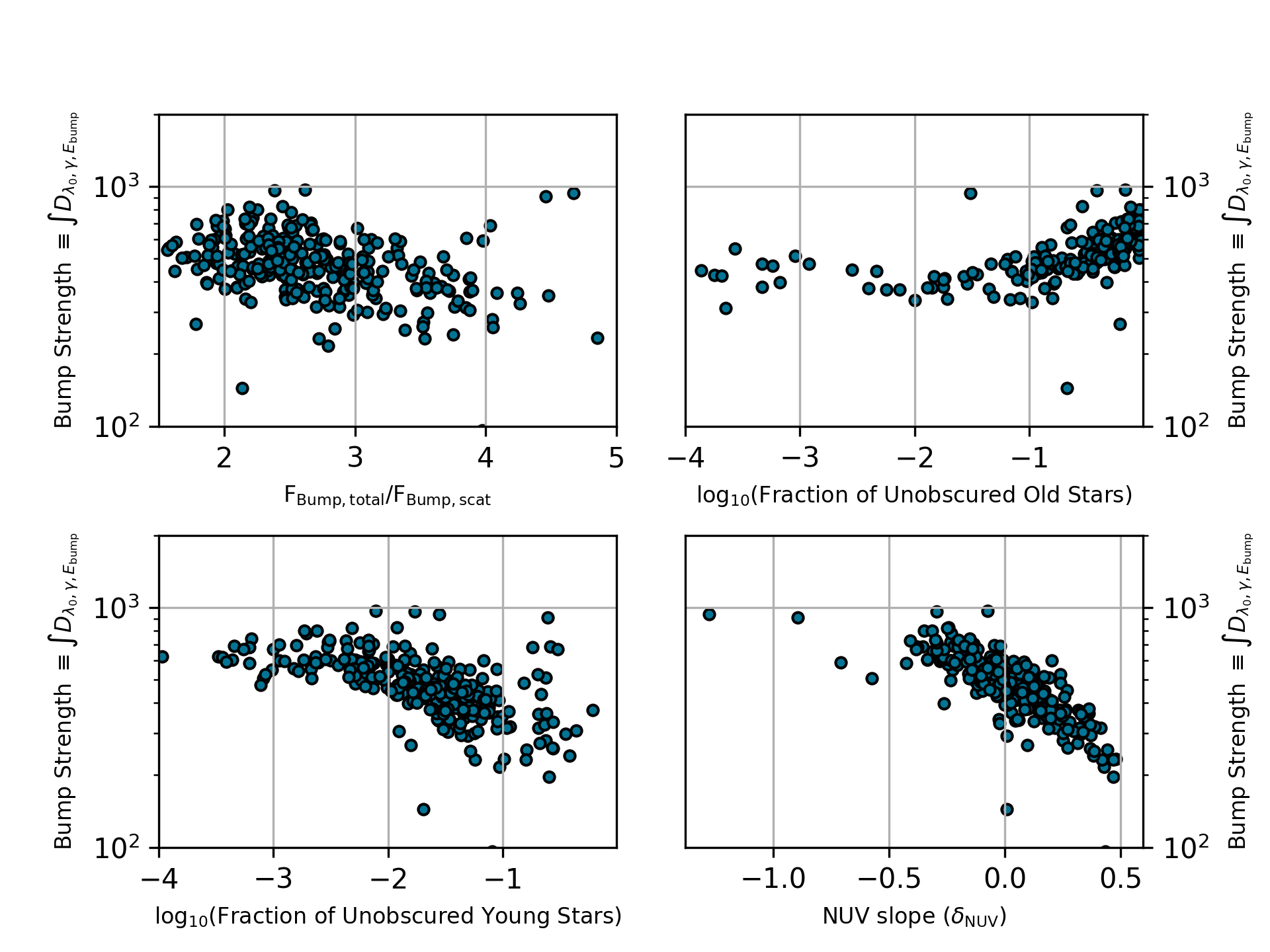}
  \end{tabular}
  \caption{Origin of variations in
    the $2175 \angstrom$ bump strength (defined as the integral of the
    best fit Drude profile across the bump).  In short, reductions of
    the bump strength are principally dictated by the fraction of
    young stars that have relatively little obscuration, and thereby
    fill in the attenuation bump.  There is a modest impact from the
    contribution of scattered light.  {\bf Top Left:} bump strength
    vs.  fraction of total light at $2175 \angstrom$ that comes from
    scattered light.  There is a weak correlation, suggesting a
    relatively small fraction of $2175 \angstrom$ absorption bumps are
    filled in by scattered light.  {\bf Top Right:} bump strength
    vs. fraction of unobscured old stars.  There is little
    correlation, due to the relatively small amount of flux in the UV
    originating from old stars.  {\bf Bottom Left:} bump strength vs
    fraction of unobscured young star emission.  As the fraction of
    naked young stars increases, the bump strength decreases.  This
    effect owes to UV emission from young stars filling in the $2175
    \angstrom$ absorption feature in the galaxy's total attenuation
    curve.  {\bf Bottom Right:} bump strength vs. NUV slope
    ($\delta_{\rm NUV}$).  Because there is a relationship between the fraction
    of unobscured young stars and the slope of the attenuation curve
    (c.f. top left panel of Figure~\ref{figure:tilt_multiplot}), there
    is a natural relationship between the $2175 \angstrom$ bump
    strength and the NUV attenuation curve slope.\label{figure:ebump_multiplot}}
\end{figure}

In Figure~\ref{figure:ebump_multiplot}, we show how the bump strength
varies in our models with a number of relevant quantities.  The top
left panel of Figure~\ref{figure:ebump_multiplot} shows the
relationship between the $2175 \ \angstrom$ UV bump strength and the
fraction of scattered light contributing to the total $2175
\ \angstrom$ flux.  There is a relatively weak correlation between the
two: light scattered into the line of sight contributes modestly to
reduced bump strengths in some attenuation curves, but it clearly does
not dominate.

In the top right and bottom left panels of
Figure~\ref{figure:ebump_multiplot}, we investigate the role of
unobscured stars in contributing to reduced bump strengths.  The top
right panel shows the bump strengths as a function of the fraction of
unobscured old ($t_{\rm age} > 50 $ Myr) stars, while the bottom left
shows the bump strengths as a function of the fraction of unobscured
young ($t_{\rm age} < 50 $ Myr) stars.  As is clear, the old stars,
which put out relatively little flux in the NUV, do not contribute
significantly to filling in the $2175 \ \angstrom$ absorption bump,
even at relatively large unobscured fractions (and in fact trend in
the opposite direction).  In contrast, however, there is a clear
relationship between the fraction of unobscured young stars, and the
UV bump strength.  In galaxies with increasingly complex young
star-dust geometries, as more young stars find low-$\tau$ UV
sightlines, this radiation fills in the UV bump and reduces its
strength.  In short: galaxies with very complex young star-dust
geometries have reduced $2175 \ \angstrom$ UV bumps.

Drawing on what we learned in \S~\ref{section:slope}, the fraction of
unobscured young stars in our simulations also correlates with the
slope of the dust attenuation curve.  It follows transitively, then,
that the bump strength should vary inversely with the attenuation
curve slope.  We quantify this relationship in the bottom right panel
of Figure~\ref{figure:ebump_multiplot}, where we show the attenuation
curve bump strength vs. NUV slope ($\delta_{\rm NUV}$;
c.f. Equation~\ref{equation:nuv}).  The greyest attenuation curves
represent galaxies with the most complex young star-dust geometry, and
therefore have the weakest bump strengths.  While we will return to
this issue further in \S~\ref{section:discussion}, but it's worth
briefly noting that \citet{kriek13a} demonstrated that for $z \sim 2$
galaxies in the NEWFIRM survey, dust attenuation slope varies
inversely with the measured bump strength.  This may provide some
tentative evidence for our interpretation.

\section{Discussion}
\label{section:discussion}
\subsection{Does a Single Attenuation Prescription Apply?}
Thus far, we have explored the physical underpinnings of variations in
dust attenuation curves in galaxies.  We now ask the slightly more
practical question: what range of attenuation curves can one expect
for galaxies at a given redshift?

To answer this, we employ the $25$ Mpc$^3$ {\sc mufasa} cosmological
hydrodynamic galaxy formation simulation \citep{dave16a}.  This
simulation has identical physics as our zoom models, save for the
inclusion of the standard {\sc mufasa} heuristic quenching model.  The
simulation is run with $512^3$ particles, but in a volume half the
size of the one our zooms are selected from, resulting in an effective
mass resolution a factor $8$ worse (i.e. larger particle masses) than
our zoom models.  Achieving the full resolution of our zoom
simulations in a cosmological volume requires a computational effort
outside the scope of the current investigation.

\begin{figure*}
  \begin{tabular}{cc}
    \hspace{-0.25cm}
    \includegraphics[width=\columnwidth]{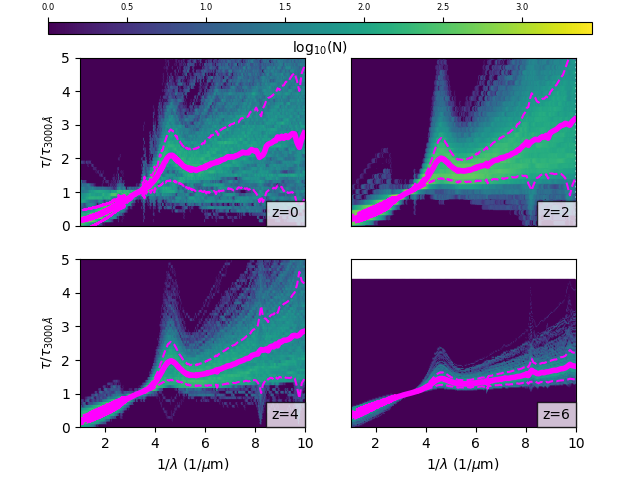}
  \end{tabular}
  \caption{\label{figure:median_curve_redshift} Heat map of
    attenuation curves at redshifts $z=0,2,4,6$ derived from {\sc
      mufasa} $25 $Mpc$^3$ cosmological simulation.  The physics of
    this simulation is identical to those in our zooms.  The solid,
    thick pink line is the median attenuation curve in all panels,
    while the dashed pink lines show the $1\sigma$ standard deviation.
    The median attenuation curves become greyer with redshift as
    geometries become more complex and median stellar ages become more
    uniform.  There is significant dispersion at all redshifts,
      though the median is typically bounded by the family of curves
      that describe local galaxies
      (c.f. Figure~\ref{figure:median_curve_redshift_literature}).  We
      publish the best fit to our median curves at all integer
      redshifts between $z=0-6$ in public repository: see text for
      details.}
\end{figure*}

\begin{figure*}
  \begin{tabular}{cc}
    \hspace{-0.25cm}
    \includegraphics[width=\columnwidth]{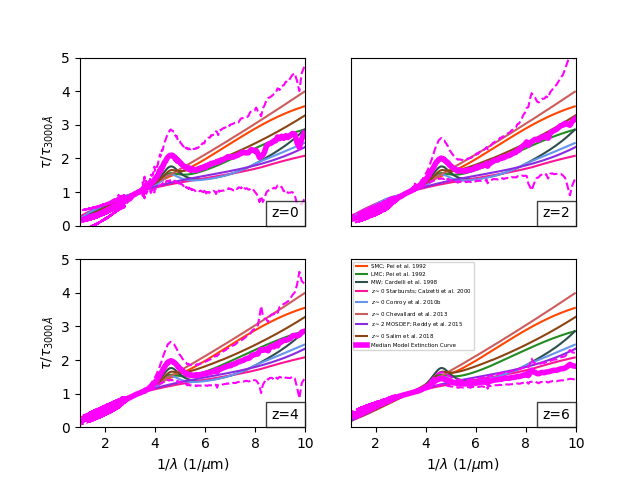}
  \end{tabular}
\caption{Median attenuation curves as a function of redshift as taken
  from Figure~\ref{figure:median_curve_redshift}, with standard
  literature curves presented for comparison.  The median curves are
  comparable to many locally calibrated curves through $z=4$, though
  become significantly greyer toward $z \sim
  6$.  See text for details. \label{figure:median_curve_redshift_literature}}\end{figure*}

In Figures~\ref{figure:median_curve_redshift}
and~\ref{figure:median_curve_redshift_literature}, we show heat maps
of the attenuation curves for all galaxies\footnote{With {\sc caesar},
  we identify galaxies as FOF groups with at least $32$ star
  particles.  This corresponds to a minimum stellar mass of $M_* > 7.2
  \times 10^{7} M_\odot$.} in the $25$ Mpc$^3$ volume at redshifts
$z=0,2,4,6$.  We additionally show the median attenuation curve
(computed by calculating the median $\tau/\tau_{3000}$ at every
wavelength $x = 1/\lambda$) at each redshift via the solid pink line,
and the $1\sigma$ standard deviation in the dashed pink lines.  In
order to best compare with observations which typically only select
star-forming galaxies, we restrict our analysis here to galaxies with
SFR$\geq 1$ $M_\odot$ yr$^{-1}$.
Figure~\ref{figure:median_curve_redshift} shows the heat map and
median attenuation curves, and
Figure~\ref{figure:median_curve_redshift_literature} shows the median
curves in comparison to literature references.

At each redshift, there is significant dispersion about the median, as
evidenced by the heat map.  In the normalized attenuation curves, a
wide range of curve tilts and bump strengths are exhibited.  The
significant dispersion seen in
Figure~\ref{figure:median_curve_redshift} is expected from
observational constraints that show a diverse range of slopes in
attenuation curves \citep[e.g.][]{wild11a,battisti17a}.  This
dispersion in curves decreases with increasing redshift.  While galaxy
geometry remains complex at these epochs, the metal content (and
hence, manifestly in these simulations, dust content) decreases, thus
reducing the impact of star-dust geometry on the attenuation curve
shape.  Beyond this, there is a much narrower distribution in
  median stellar ages with increasing redshift (c.f. \S~\ref{section:slope}).

Despite the strong dispersion seen at nearly all redshifts, the
medians in the broad distribution of modeled attenuation curves
are always bounded by the standard literature attenuation
  curves.  This is demonstrated in
  Figure~\ref{figure:median_curve_redshift}, where we compare against
  standard literature curves.  For convenience, we have published the
best fitting median curves in
Figure~\ref{figure:median_curve_redshift} at integer redshifts from
$z=0-6$ on a publicly accessible
website\footnote{\url{https://bitbucket.org/desika/narayanan_attenuation_laws/}}.
It is important to note that while the median values are appropriate
for ensemble averages, these median values (or any assumption of a
locally-calibrated curve, for that matter) may grossly mischaracterize
the underlying attenuation on a case by case basis.

Finally, it is worth noting that the most common attenuation curves in
any of our modeled redshift bins all have prominent bump features.  It
is indeed possible to generate attenuation curves with relatively
small bump contributions: this is demonstrated explicitly in
Figures~\ref{figure:all_curves} and~\ref{figure:ebump_multiplot}.
These curves with minimal bumps owe their origin solely to geometry and
radiative transfer effects (i.e. no modification of the underlying
dust properties is necessary).  This said, this is not typical in our
models: the median curve within any redshift bin in
Figure~\ref{figure:median_curve_redshift} displays a prominent bump.

\subsection{Comparison to Other Models and Constraints}
To our knowledge, this paper represents the first cosmological
hydrodynamic simulation of galaxy formation investigating theoretical
dust attenuation curves in galaxies.  This said, our work builds on a
deep theoretical literature.  In this section, we aim to place the
results from our work in this context.  We painted the landscape for
theoretical work in this field in \S~\ref{section:introduction}.  In
this section, we aggregate some key results from these papers, and
compare them to our own investigation.

\subsubsection{On the Role of Geometry}
In Figures~\ref{figure:frac_nodust}, and~\ref{figure:tilt_multiplot},
we demonstrated via both population synthesis experiments as well as
direct cosmological simulation that the role of the star-dust geometry
is paramount in driving the tilt of normalized attenuation curves.
For young galaxies, an increasing fraction of unobscured young stars
flattens normalized attenuation curves, while for galaxies dominated
by older stellar populations, an increasing fraction of unobscured old stars
steepens normalized curves.

In general, at least the former point (and more broadly, the idea that
geometry plays an important role in setting the shape of attenuation
curves), is already well-appreciated in the theoretical literature.
Indeed, a broad range of modeling techniques arrive at the same
conclusion.  For example, \citet{witt96a} examined radiative transfer
models in a two-phase clumpy medium, and found that as a general rule,
attenuation curves became greyer (flatter) as the obscuration
inhomogeneity increased.  \citet{seon16a} expanded upon these results
by using a turbulent studying the radiative transfer in a turbulent
medium, and arrived at a similar conclusion, while \citet{natale15a}
utilized a coupling of 3D dust radiative transfer calculations (as in
this paper) with idealized hydrodynamic models of disks in evolution
to demonstrate this principle.  \citet{fischera11a} additionally
employed non-homogeneous ISM models to explain grey attenuation curves.

It is important to note that the concept of increased complexity in
the star-dust geometry driving greyer attenuation curves only applies
to galaxies whose light are dominated by young stellar populations
(c.f. Figure~\ref{figure:tilt_multiplot}).  In our model, galaxies
whose luminosity is dominated by older stellar populations have
normalized attenuation curves that become {\it steeper} as more old
stars are decoupled from dust.

\subsubsection{Bump Strengths}
When assuming an attenuation curve for the purposes of SED fitting in
low-metallicity galaxies (especially at high-redshift), a common
assumption is an SMC-like attenuation curve.  This is motivated by the
assumption that the bumpless curve from the SMC may owe its origin to
different grain compositions (presumably driven by the galaxy's low
metallicity) than in the Milky Way.  Similar logic oftentimes motivate
the usage of a bumpless \citet{calzetti00a} curve to observations of
heavily star-forming galaxies.

Throughout this work, we have demonstrated that even without changing
the underlying grain properties (i.e., our extinction curve in every
model is identical), we are able to generate extinction curves that
have a diverse range in bump strengths (see,
e.g. Figure~\ref{figure:median_curve_redshift} for one such example).
In other words, curves with very small bump strengths in our models
result simply from the geometry of the system.  This conclusion is not
necessarily shared amongst other theoretical models.

For example, \citet{witt00a} expanded on the radiative transfer models
of \citet{witt96a}, and found that they required a change to the
intrinsic dust curve in order to make a bumpless curve.  In specific,
only by employing an extinction curve without a $2175 \angstrom$ bump
were they able to reproduce observed bump-free curves.  Others
\citep[e.g.][]{fischera11a} have suggested models in which the UV bump
carriers are destroyed at threshold column densities in order to
reproduce the \citet{calzetti00a} curve.

\citet{hou17a} implemented a full dust formation and destruction model
into cosmological hydrodynamic simulations, and track the grain size
distribution and extinction curve.  While they do not model the
radiative transfer associated with these simulations, they find in the
{\it extinction} curve that $2175 \angstrom$ bump strengths are
naturally correlated with dust growth dominated by accretion of metals
onto dust grains.  While this study does not explore the geometry
effects that drive {\it attenuation} curves, it is certainly
conceivable that intrinsic dust properties can impact the strength of
the UV bump.

At the same time, other models have suggested, like this work, that it may be possible to achieve 
bump-free attenuation curves without
modifying the underlying extinction curve.  For example,
\citet{granato00a} developed semi-analytic models to demonstrate that
the bump-free \citet{calzetti00a} curve may be a result of a 'birth
cloud' model, wherein the radiation from young stars are heavily
attenuated by their birth clouds, but older stars are only attenuated
by diffuse ISM.  In this picture, the observed UV is dominated by an
older stellar population which sees relatively little diffuse dust,
and hence the emission from these objects can fill in the $2175
\angstrom$ absorption feature.  This semi-analytic model was expanded
upon by \citet{panuzzo07a}, who found similar results.

While the \citet{granato00a} and \citet{panuzzo07a} studies share in
common with our model the idea that an attenuation curve with very
small bump strengths is achievable without changing the underlying
extinction properties of grains, the models share a key difference.
\citet{granato00a} and \citet{panuzzo07a} find that as young stellar
populations become more obscured, $2175 \angstrom$ bump sizes decrease
in attenuation curves.  In contrast, in
Figure~\ref{figure:ebump_multiplot}, we demonstrate that the opposite
is true for our models: as a larger fraction of young stars becomes
unobscured, observed bump strengths decrease owing to the UV radiation
from these young stars filling in the $2175 \angstrom$ absorption
feature.  An important corollary to our model is that the same
displacement between O and B stars and sites of dust obscuration
result in greyer (flatter) attenuation curves, leading to a natural
relationship where shallower attenuation curves have smaller bump
features in our model (bottom right panel of
Figure~\ref{figure:ebump_multiplot}).

\citet{seon16a} similarly derived a model for dust attenuation curves
that exhibits a relationship between bump strength and slope of the
attenuation curve in the same direction as both our models, and the
\citet{kriek13a} observations.  This too can be attributed to geometry
effects, wherein as the clumping or size of the source distribution
increase, attenuation curves become greyer and exhibit smaller $2175
\angstrom$ bumps.  We note, however, that a \citet{calzetti00a} curve
is only attainable using the \citet{weingartner01a} Milky Way dust
model when the intrinsic bump is removed or suppressed in the \citet{seon16a} model.  As a result,
their model may be viewed as intermediate between the results of
\citet{granato00a,panuzzo07a} and ours.

\begin{figure}
  \begin{tabular}{cc}
    \hspace{-0.25cm}
    \includegraphics[width=\columnwidth]{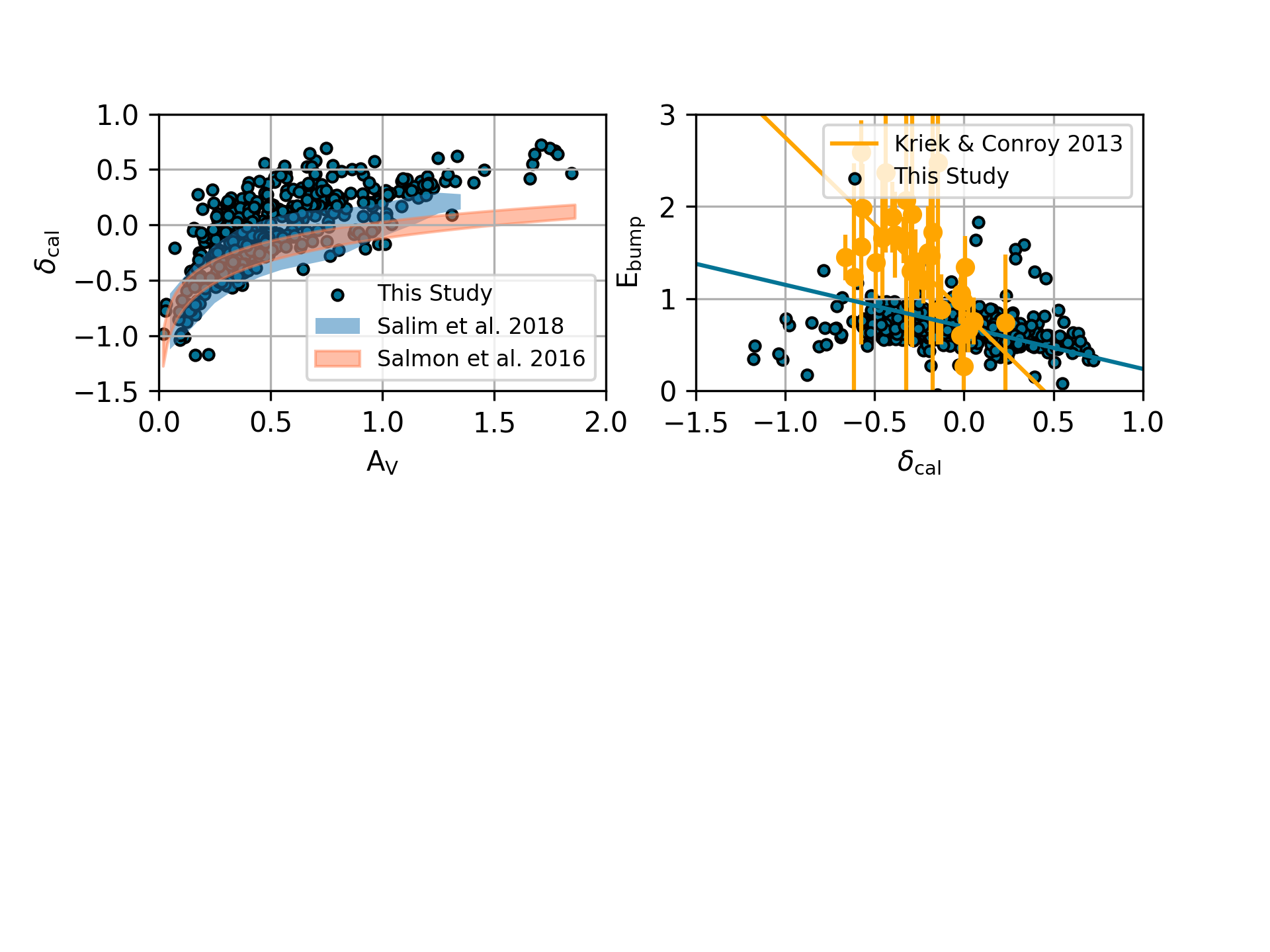}
  \end{tabular}
  \vspace{-7cm} 
  \caption{Comparison with observations.  Left: power-law index of
    curve versus $V$-band extinction.  Blue and orange shaded
    regions denote observations, while blue points show model zoom
    simulations.  Right: UV bump normalization (assuming a fixed
    width) against powerlaw index.  See \S~\ref{section:observational_comparison} for details on specific comparisons.\label{figure:salim} }
\end{figure}

\subsection{Comparison with Observations}
\label{section:observational_comparison}
We now turn to a comparison of our models with relatively recent
  observational results in this area.  We remind the reader of the
  discussion surrounding Equations~\ref{equation:modified_calzetti}-\ref{equation:ebumpkc} and
  Table~\ref{table:fit}.  In specific, while we find a piece-wise fit
  to provide the best fits to our model attenuation curves as outlined
  in \S~\ref{section:fitting}, many observational studies employ a
  modified \citet{calzetti00a} relation, where both a Drude-like
  profile for the UV bump as well as a power-law modification are
  employed \citep[e.g.][]{noll09a}.  As we clarify in
  Table~\ref{table:fit}, we distinguish the power-law index derived
  from this method of fitting ($\delta_{\rm cal}$) from the power-law
  we typically employ just in the NUV bands ($\delta_{\rm NUV}$).  

We compare to the observational results of \citet{kriek13a},
\citet{salmon16a} and \citet{salim18a} in Figure~\ref{figure:salim}.
\citet{salmon16a} and \citet{salim18a} derive attenuation laws for
redshift $z\sim 2$ and $z\sim 0$ galaxies, respectively, via SED
fitting techniques.  Evident from both of these studies is a
relationship between the $V$-band optical depth and the slope of the
attenuation curve, $\delta_{\rm cal}$.  This is similar to
the powerlaw relationship modeled by \citet{chevallard13a} and
\citet{leja17a} between the optical depth of diffuse dust and the
powerlaw slope of the attenuation curve.  In the left panel of
Figure~\ref{figure:salim}, we show a comparison between our models and
this observed trend.  We include every galaxy in our sample of zooms,
though note that the \citet{salmon16a} study and \citet{salim18a}
study of course both employ individual selection techniques within
particular redshift ranges.  Given both this, as well as the relative
uncertainties involved in deriving attenuation curves from SED
fitting, the trend of increasing $\delta_{\rm cal}$ with $V$-band
optical depth in the model galaxies is encouraging.

\citet{kriek13a} employed SED fitting techniques to observations
  of $z\sim 2$ galaxies to derive a relationship between the UV bump
  strength and slope power-law index, $\delta_{\rm cal}$.  In
  Figure~\ref{figure:ebump_multiplot}, we demonstrated a similar
  relationship, though characterized this in terms of the integrated
  Drude profile, $\int D_{\lambda_0,\gamma,E_{\rm bump}}$, and the NUV
  slope, $\delta_{\rm NUV}$.  \citet{kriek13a} fix the width of the
  bump to $\gamma_{\rm const} = 350 \angstrom$, and therefore characterize the strength of
  the bump by its normalization, $E_{\rm bump,KC}$
  (c.f. Table~\ref{table:fit}).  In order to best compare to this
  study, we have re-performed our fits by fixing our bump widths
  similarly, and report in the right side of Figure~\ref{figure:salim}
  our modeled relationship between $E_{\rm bump}$ and $\delta_{\rm
    cal}$.  We show our model points in blue, and compare these to the
  \citet{kriek13a} data in orange.  The best fit lines for both are
  shown.  Our model galaxies show a similar trend as the
  \citet{kriek13a} observations in that steeper curves tend to have
  more prominent bump strengths.  In our model, this owes primarily to
  unobscured young stars reducing bump strengths.  Our modeled best
  fit relation is: 
\begin{equation}
  \label{equation:ebumpkc_fit}
  E_{\rm bump,KC} = -0.46 \times \delta_{\rm cal} + 0.69
\end{equation}

  Our model galaxies exhibit a much shallower gradient than
    what is observed.  This may owe to a number of causes.  First, our
    modeled bump widths, $\gamma$ (c.f. Equation~\ref{equation:nuv})
    span a broad range of values, with some widths exceeding twice the
    assumed $\gamma_{\rm const} = 350 \angstrom$ employed for the construction of
    Figure~\ref{figure:salim}.  Forcing a bump strength, therefore,
    may result in poorly performing fits, and therefore $E_{\rm bump}$
    values that do not reflect the true dynamic range of bump
    strengths.  As an example, examination of the bottom right panel
    of Figure~\ref{figure:ebump_multiplot} demonstrates that when
    considering the integral of the Drude profile of the bump
    strength, we see a dynamic range spanning an order of magnitude in
    bump strength, unlike the factor $\sim 2-4$ when characterizing
    the bump strength by $E_{\rm bump}$ alone.  Beyond this, a true
    apples-to-apples comparison between our model points and the
    observed data would involve our fitting our model SEDs (assuming a
    given star formation history and IMF), and recovering the inferred
    attenuation law accordingly.  Exploring the differences resulting
    in inferred attenuation curves from SED modeling from those
    directly modeled will be presented in future work.  We
    note that \citet{seon16a} derived relationships between $E_{\rm bump,KC}$ and
    $\delta_{\rm cal}$ that were typically steeper than the
    \citet{kriek13a} relation.  More observational and theoretical
    work in this area is warranted.


\section{Conclusions and Summary}
\label{section:summary}
We have developed a model for the origin of variations in dust
  attenuation curve shapes and bump strengths.  We accomplish this by
  combining high-resolution cosmological galaxy formation simulations
  with 3D dust radiative transfer calculations.  Critically, for these
  radiative transfer models, we {\it hold the underlying extinction
    curve fixed}, and ask how geometry and radiative transfer effects
  impact the resultant attenuation curves. Our main results follow:
\begin{enumerate}
  \item Despite the usage of a constant extinction curve in our
    underlying radiative transfer calculations, we find dramatic
    variations in the derived attenuation laws.  These variations
    depend primarily on complexities in the star-dust geometry
    (Figures~\ref{figure:frac_nodust},~\ref{figure:all_curves},
    \&~\ref{figure:tilt_multiplot}).  In detail:
    \begin{enumerate}
    \item Increasing fractions of unobscured young stars result in flatter (greyer) attenuation curves as the galaxy becomes more transparent to UV radiation (Figures~\ref{figure:frac_nodust} \&~\ref{figure:tilt_multiplot}).
    \item Increasing fractions of unobscured evolved stars results in
      steeper attenuation curves
      (Figures~\ref{figure:gergo},~\ref{figure:stellar_ages_sfh} \&~\ref{figure:tilt_multiplot}).
    \item These results taken together drive a trend where galaxies
      with highly obscured sightlines toward young stars, but a
      significant (unobscured) evolved stellar population will have the steepest
      normalized attenuation
      curves (Figure~\ref{figure:tilt_multiplot}).
    \end{enumerate}
    \item The $2175 \angstrom$ UV bump strengths vary dramatically,
      despite our usage of an extinction curve with a UV bump present.
      Unobscured O and B stars result in reduced bump strengths in our
      model, with scattered light only having a secondary effect on
      the feature (Figure~\ref{figure:ebump_multiplot}).
      \item The combined effect of unobscured young stars both
        flattening attenuation curve slopes, as well as reducing the
        bump strength results in a natural relationship wherein the
        slope of the attenuation curve is related to the bump
        strength: flatter attenuation curves tend to have smaller bump
        strengths (Figure~\ref{figure:ebump_multiplot}).
\item We apply these results to a $25 $ Mpc/h cosmological volume to
  derive the median curve and expected dispersion at integer redshifts
  from $z=0-6$.  While the median curve at a given redshift is
  typically bounded by standard literature curves, the dispersion is
  significant.  The average dispersion decreases with increasing
  redshift, and the median curves become greyer.  This owes to reduced
  dispersion in star-dust geometry, as well as narrower distribution
  in median stellar ages with redshift
  (Figure~\ref{figure:median_curve_redshift}).  We publish these median curves on a public-facing website.
\end{enumerate}

\section*{Acknowledgements}
D.N. is grateful to Andrew Battisti, Daniela Calzetti, Rob Kennicutt,
Maciej Koprowski, Karin Sandstrom, Samir Salim, Brett Salmon and
George Privon for valuable conversations during this study.  We
additionally thank Mariska Kriek, Samir Salim and Brett Salmon for
providing data to us for our comparisons with observations. The
simulations published here were run on the University of Florida
HiPerGator supercomputing facility, and the authors acknowledge the
University of Florida Research Computing for providing computational
resources and support that have contributed to the research results
reported in this publication.  This study was funded in part by NSF
AST-1715206 and HST AR-15043.0001.


\bibliographystyle{mnras} \bibliography{/Users/desika.narayanan/Dropbox/paper/full_refs}

\end{document}